\documentclass[11pt]{article}

\usepackage[preprint]{acl}

\usepackage{times}
\usepackage{latexsym}

\usepackage[T1]{fontenc}

\usepackage[utf8]{inputenc}

\usepackage{microtype}

\usepackage{inconsolata}

\usepackage{graphicx}
\usepackage[capitalise,nameinlink,noabbrev]{cleveref}
\usepackage{booktabs}
\usepackage{multirow}
\usepackage{amssymb}
\usepackage{tcolorbox}
\usepackage{verbatim}
\usepackage{listings}
\usepackage{caption}
\usepackage{adjustbox}
\usepackage{tabularx}
\tcbuselibrary{listings,breakable}

\title{MiST: Mid-Training LLMs for Cybersecurity}

\author{
  Oded Ovadia\thanks{Equal contribution.}\quad
  Elad Ben Zaken\footnotemark[1]\quad
  Elad Guttman\footnotemark[1]\quad
  Orly Moreno Kadosh \\
  Dream \\
  \texttt{\{odedov,elad,eladg,orly\}@dreamgroup.com}
}

\begin{document}
\maketitle
\begingroup
\renewcommand{\thefootnote}{}
\footnotetext{Accepted to the Main Conference of EMNLP 2026.}
\endgroup

\begin{abstract}

    Cybersecurity combines high-stakes analysis with complex technical language, making it an impactful and challenging domain for LLMs. We present MiST (Mid-trained Security Transformer), a suite of 8B and 32B models that achieve strong performance on public cybersecurity benchmarks. We use mid-training as an intermediate adaptation stage between general pre-training and cybersecurity training. Rather than performing continual pre-training over large volumes of raw domain text, we curate a compact, expert-vetted seed corpus, and transform it into high-quality domain-specific synthetic training data. The final MiST checkpoints improve mean cybersecurity accuracy by +13.1 and +8.6 absolute percentage points over the corresponding Qwen baselines for 8B and 32B, respectively, corresponding to relative gains of +27.0\% and +15.8\%. Ablation results further show that these cybersecurity gains arise in the mid-training and supervised fine-tuning stages through a combination of the synthetic data generation flows. Furthermore, we show that MiST provides a stronger initialization for downstream task-specific fine-tuning adaptation and reinforcement learning. 
    
\end{abstract}

\section{Introduction}
\begin{figure*}[!htb]
    \centering
    \includegraphics[width=1.0\linewidth]{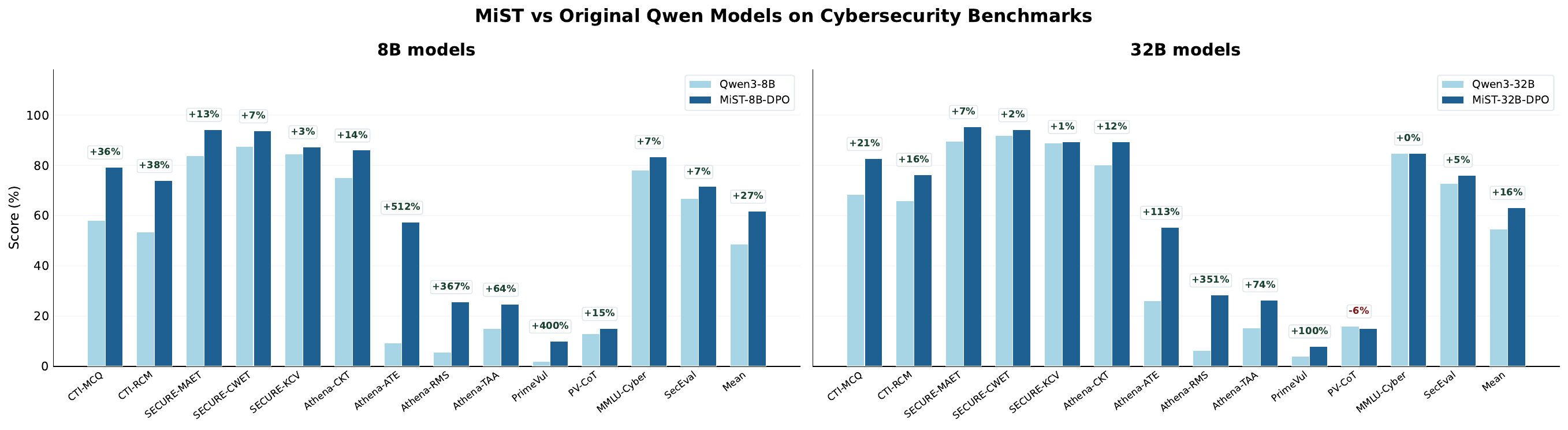}
    \caption{Accuracy of MiST and corresponding Qwen models on cybersecurity benchmarks, shown for 8B (left) and 32B (right), with relative improvements annotated. Results are reported relative to the original Qwen baselines from which MiST is derived.}
    \label{fig:improvement}
\end{figure*}

\paragraph{LLMs and Cybersecurity.} As large language models (LLMs) continue to improve on
natural-language and code tasks~\citep{minaee2024large,jiang2024survey}, interest has grown
in applying them to cybersecurity~\citep{xu2024large,zhang2025llms}. 
Security analysis requires interpreting vast amounts of textual data, such as
vulnerability records, attack taxonomies, code, and reports, making LLMs a promising
tool for supporting cybersecurity analysis.
As cybersecurity threats continue to increase in scale and sophistication~\citep{li2021comprehensive},
there is a growing need for models that can operate over such data.


At the same time, cybersecurity exposes key limitations of general-purpose LLMs.
Effective performance in this domain demands precise interpretation of specialized terminology, dynamic attack techniques, and context-dependent operational knowledge. These are requirements that general-purpose training does not necessarily meet, especially as high-quality cybersecurity pre-training corpora remain scarce~\citep{yu-etal-2025-primus}.
Furthermore, privacy-sensitive environments typically mandate on-premises deployment, making reliance on external API-based models impractical~\citep{huang2025middle}.

\paragraph{Domain-specific models.} One approach to such challenges is to develop domain-specific models tailored to particular
areas. By incorporating domain data, terminology, and expert knowledge, such models aim to
better capture specialized semantics and knowledge than general-purpose
LLMs~\citep{ling2023domain,wang2023survey}. Prior work has demonstrated the effectiveness
of this strategy in domains such as healthcare~\citep{wu2024pmc},
finance~\citep{bhatia2024fintral}, and software
engineering~\citep{hui2024qwen2,guo2024deepseek,huang2025opencoder}. These efforts commonly
follow a two-stage methodology: continual pre-training (CPT) on large-scale raw domain
text, followed by supervised fine-tuning (SFT) on instruction data.

Mid-training provides a more deliberate alternative to this raw-domain adaptation pipeline.
In this work, we use mid-training as a domain-specific stage after pre-training and before post-training, 
designed to bridge the distributional gap between a broad base model and the target domain~\citep{tu2025survey,zhang2025interplay}.
In contrast to raw CPT, which primarily focuses on scaling the number of new domain tokens, cybersecurity mid-training in this work emphasizes the composition of the training data: we start from a small, high-quality seed dataset and extend it with diverse synthetic pipelines to form our mid-training corpus, thereby facilitating better domain learning.

\begin{figure*}[!hbt]
    \centering
    \includegraphics[width=1.0\linewidth]{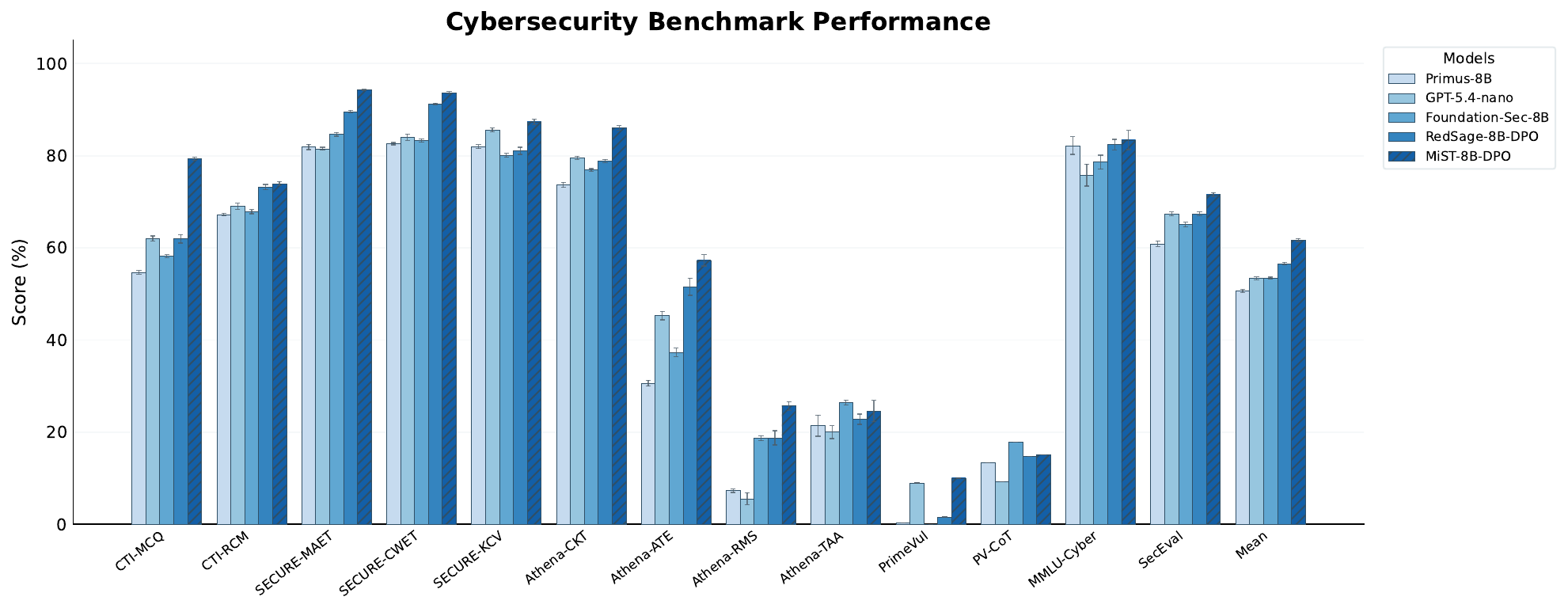}
    \caption{Performance on the cybersecurity evaluation suite (\Cref{sec:cyber_benchmarks}), comparing MiST-8B with representative general and cybersecurity models of similar size. Bars report accuracy (\%) as reported in \Cref{tab:cyber_results}, along with standard deviation error bars. This comparison is particularly relevant as cybersecurity models often need to be deployed in resource-constrained environments, where strong performance at smaller model sizes is especially valuable. MiST-8B greatly outperforms other models in its weight class.}
\label{fig:cyber_benchmarks}
\end{figure*}

\paragraph{MiST: Cybersecurity Language Model.} We present MiST, a family of 8B and 32B models adapted from Qwen checkpoints. Our
approach centers on the construction of the mid-training corpus. We curate a compact,
expert-vetted seed corpus and use it to generate synthetic training data that teaches
cybersecurity concepts, terminology, and structured analysis. 
We first mid-train on this corpus, then perform SFT using a combination of general and cybersecurity instruction data, and finally apply preference optimization as a general alignment stage. Since this final DPO stage uses general preference data rather than cybersecurity-specific preferences, we do not treat it as the main source of cybersecurity capability; instead, it is intended to preserve the domain capabilities acquired during mid-training and SFT while improving general response behavior.
On public cybersecurity benchmarks, the MiST models substantially improve over their base models (\Cref{fig:improvement}), outperform similarly sized cybersecurity-specialized models (\Cref{fig:cyber_benchmarks}), and are competitive with larger LLMs (\Cref{tab:cyber_results}).


More broadly, MiST frames mid-training as a corpus design problem rather than a
token-scaling problem. In cybersecurity, where much of the relevant knowledge is encoded
in compact expert artifacts such as vulnerability records, taxonomies, detection rules,
and threat-intelligence reports, transforming these sources into an intermediate training
distribution can be more effective than simply training on more raw domain text.

\begin{figure*}[!htb]
    \centering
    \includegraphics[width=0.9\linewidth]{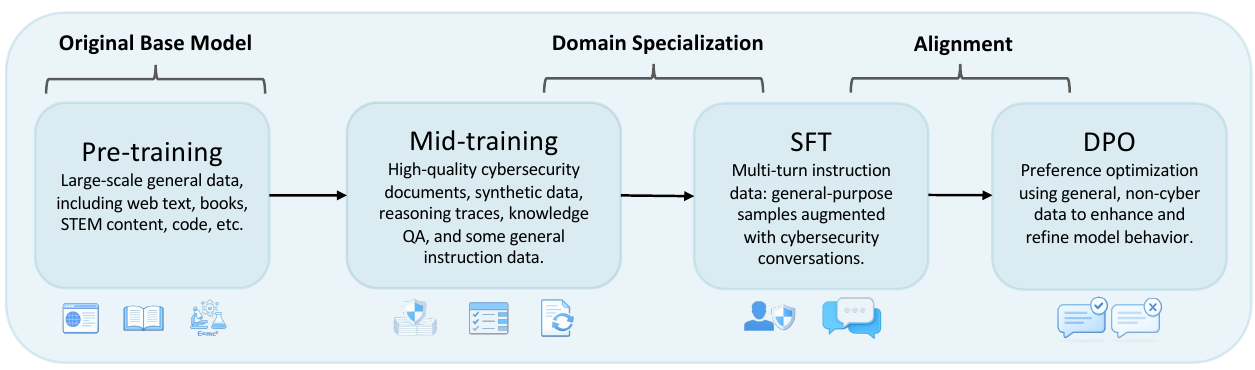}
    \caption{Overview of the full training pipeline. Training begins from a general-purpose pre-trained base model. Mid-training is then used to enhance cybersecurity-specific capabilities using high-quality data. Supervised fine-tuning (SFT) introduces the chat template and further refines general behavior while strengthening cybersecurity capabilities. Direct preference optimization (DPO)~\citep{rafailov2023direct} is applied in the final stage to improve general alignment.}
    \label{fig:midtrain_flow}
\end{figure*}

\begin{figure*}[!htb]
    \centering
    \includegraphics[width=0.85\linewidth]{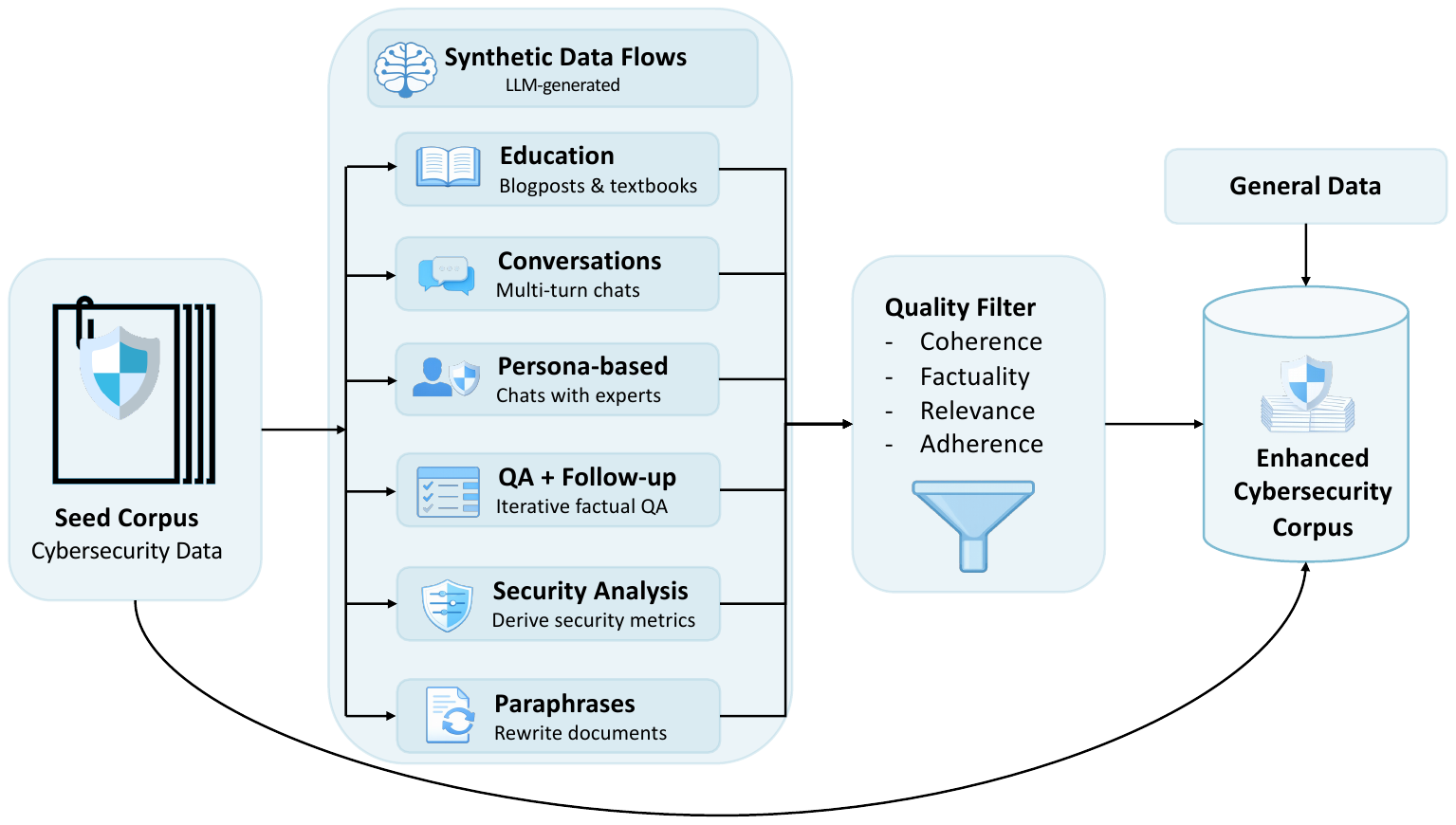}
    \caption{Overview of the synthetic data generation pipeline. An expert-curated cybersecurity seed corpus (\Cref{sec:seed}) is expanded through multiple data generation flows, producing synthetic educational content, multi-turn conversations, persona-based interactions, iterative QA, and analytical explanations (\Cref{sec:synth}). All generated data points are filtered for coherence, factuality, relevance, and instruction adherence using an LLM, and then combined with a small amount of general SFT data to form the final cyber mid-training corpus. Detailed examples of generated samples are provided in \Cref{app:examples}.}
    \label{fig:cyber_flow}
\end{figure*}

\section{Related Work}

 \paragraph{Cybersecurity LLMs}
Prior work has explored several strategies for adapting general-purpose LLMs to
cybersecurity. PRIMUS~\citep{yu-etal-2025-primus} and
Foundation-Sec-8B~\citep{kassianik2025llama,weerawardhena2025llama} emphasize continued
pretraining (CPT) on large raw cybersecurity corpora, while community models such as
DeepHat-V1~\citep{deephatv1} and Lily-Cybersecurity~\citep{lilycybersecurity} primarily
report cybersecurity-focused SFT through model cards. A more data-centric line constructs
cybersecurity supervision from expert sources: CyberPal.AI~\citep{levi2025cyberpal}
introduces SecKnowledge, an expert-guided instruction dataset expanded with
content-grounded synthetic generation. Recent and contemporaneous works use synthetic
data at other stages: CyberPal 2.0~\citep{levi2025toward} enriches cybersecurity
instructions with grounded reasoning traces, while RedSage~\citep{suryanto2026redsage}
combines large-scale cybersecurity CPT with agentically generated multi-turn SFT data. MiST
instead uses synthetic transformations of a compact expert-vetted corpus as the
mid-training data itself, exposing the base model to high-quality synthetic cybersecurity data before 
proceeding to SFT and preference tuning.

\paragraph{Mid-training} Mid-training has recently emerged as a distinct stage in the LLM training pipeline. In a
detailed review of the topic, \citet{tu2025survey} describe mid-training as ``the critical
bridge between general pre-training and post-training.'' A useful way to make this bridge
concrete is through the data distribution. \citet{zhang2025interplay} argue that
mid-training works by moving the model toward the target post-training distribution before
post-training begins, with the strongest gains when the intermediate data is closer to the
target domain than ordinary general pre-training data. This view is also consistent with
evidence from parameter-efficient fine-tuning that adaptation can often expose or redirect
knowledge already present in pretrained models, rather than requiring all task-relevant
knowledge to be learned from scratch~\citep{benzaken2022bitfit}. Under this view, mid-training is not defined
only by its stage in the training pipeline, but also by the construction of the intermediate corpus.

 This paradigm has been adopted in several recent large-scale efforts~\citep{wang2025octothinker,abdin2024phi3technicalreporthighly,wake2024yi,olmo2025olmo}. Notably, \citet{olmo2025olmo} devote a substantial portion of their training methodology to mid-training, constructing a high-quality corpus of approximately 100B tokens. Their mid-training data combines newly generated synthetic sources with carefully filtered and rewritten data derived from the pre-training stage, explicitly tailored to the target capabilities emphasized at this phase. This differs from raw CPT, where the model is typically trained further on domain text with little change to its original format. In our setting, the same distinction motivates a synthetic cybersecurity corpus built from authoritative sources but rewritten into forms that expose the model to various aspects of the concepts, relations, and reasoning used in security applications.

\section{Methodology}\label{sec:cyber_midtraining_data}
In this section, we describe our adaptation strategy as outlined in
\Cref{fig:midtrain_flow}, with a particular emphasis on the synthetic data generation
methodology visualized in \Cref{fig:cyber_flow}.

 We first curate a small but information-dense seed corpus composed of high-quality cybersecurity documents (\Cref{sec:seed}). Rather than use this corpus directly as raw CPT data, we treat it as source material for a synthetic data generation pipeline tailored to cybersecurity tasks. The resulting data emphasizes knowledge acquisition, concise reasoning, and multi-turn conversational behavior, with rigorous quality verification (\Cref{sec:synth}). Finally, we discuss the use of non-cybersecurity data to preserve general capabilities (\Cref{sec:general_mix}) and describe the training process (\Cref{sec:training}). Full dataset composition and statistics across the training stages are reported in \Cref{tab:data_stats}.



\subsection{Seed Data Curation}\label{sec:seed}

Our seed corpus exposes the model to complementary forms of cybersecurity knowledge rather than a single homogeneous text distribution. We organize it into four source families: vulnerability and threat intelligence; security knowledge bases and taxonomies; operational security artifacts; and defensive guidance and platform documentation. Because these sources vary in structure, granularity, audience, and density, we apply source-specific transformations that convert structured or specialized artifacts into examples better suited for cybersecurity reasoning. Brief descriptions of the cybersecurity resources, taxonomies, and acronyms referenced in this section are provided in \Cref{app:cyber_resources}.

\paragraph{Vulnerability and threat intelligence.}
We include NVD CVE records along with expert curated CTI/RSS reports. These sources provide instance-level knowledge about real vulnerabilities and threats, including identifiers, advisory text, CVSS metadata, affected products, CWE mappings, references, and vulnerable or patched implementations. Since many records are brief or operational, we use them as anchors for synthetic transformations covering severity, impact, remediation rationale, weakness mappings, and mitigation intuition.

\paragraph{Security knowledge bases and taxonomies.}
We use CWE, CAPEC, ATT\&CK, and D3FEND to capture cybersecurity abstractions and relationships. CWE covers recurring software and hardware weaknesses; CAPEC describes attack patterns; ATT\&CK organizes adversary tactics, techniques, procedures, and software; and D3FEND represents defensive countermeasures. 

\paragraph{Operational security artifacts.}
To connect abstract knowledge to practitioner workflows, we include Sigma rules, Atomic Red Team tests, Splunk ESCU detections, and MISP Galaxy clusters. These sources expose detection logic, log-source assumptions, analytic stories, adversary-emulation steps, malware and threat-actor clusters and ATT\&CK mappings. They help the model link techniques, weaknesses, and attack patterns to detection engineering, threat hunting, incident analysis, and red-team emulation.


\paragraph{Defensive guidance and platform documentation.}
We include OWASP and NIST materials, vendor and platform documentation, cloud and
container documentation, operating-system and browser sources, and a security-focused
subset of English Wikipedia. These sources provide defensive guidance, platform
terminology, APIs, configuration patterns, security mechanisms, operational constraints,
and background context. Wikipedia is used only for terminology and historical context;
advisories, standards, and vendor documentation remain the primary sources for security
facts.

\subsection{Synthetic Data Generation}\label{sec:synth}
We create multiple synthetic data generation pipelines grounded in the curated seed corpus
to strengthen cybersecurity knowledge and capabilities. We group these pipelines by
training stage: mid-training flows emphasize knowledge acquisition and foundational
reasoning, while post-training flows focus on conversational abilities for realistic
cybersecurity interactions. All flows were manually reviewed and refined, with additional
dataset curation details and representative examples provided in \Cref{app:examples}.


 \paragraph{Generation model.} We select
Qwen-30B-A3B\footnote{\url{https://huggingface.co/Qwen/Qwen3-30B-A3B}}
as the synthetic data generator based on its performance and efficiency. It provides strong
instruction-following and reasoning capabilities while activating only 3B
parameters, enabling fast and scalable inference. While it's smaller than many frontier
models, our generation setup consistently supplies context from the seed documents,
allowing the model to function primarily as a transformation engine rather than relying on
memorized knowledge.

\subsubsection{Flows for Mid-Training Data}
\paragraph{Paraphrasing.}
Training on semantically diverse rephrasings of the same content is known to improve knowledge acquisition in LLMs ~\citep{ovadia2024fine,team2025kimi,ovadia2025knowledge}. Therefore, we implement a semantic rewriting pipeline
that generates paraphrased variants of seed documents by varying lexical choices and
syntactic structure while preserving technical meaning, thereby building a linguistically
diverse view of core cybersecurity concepts.

 \paragraph{Educational transformation.} Educational-style data has been shown to be effective for training
LLMs~\citep{li2023textbooks,penedo2024fineweb}. Each seed document is transformed into a
highly structured educational artifact, written either in the style of a professional
cybersecurity blog post or as a textbook-style chapter.

 \paragraph{QA.} Beyond longer conversational scenarios, we generate concise factual question--answer pairs
that can be answered in one or two sentences, encouraging accurate and succinct responses.
To improve coverage, we iteratively extend this process by reusing previously generated
questions as context for follow-up question generation, explicitly instructing the model to
target aspects not addressed in earlier rounds; we perform two additional follow-up rounds.

 \paragraph{Cyber metrics and terminology analysis.} For documents that reference structured security metadata (e.g., CVE/CWE identifiers, CVSS
vectors and scores, or ATT\&CK techniques), we provide the relevant fields and prompt the
model to explain and justify them. This emphasizes domain reasoning while avoiding
error-prone, unguided generation of security metadata.

\subsubsection{Flows for Post-Training Data}
\paragraph{Conversations.} We generate full user--assistant dialogues in which the assistant acts as a helpful
cybersecurity expert and the user queries are also model-generated. Conversations run for
up to seven turns, but may terminate earlier when they reach a natural stopping point.

 \paragraph{Persona-based conversations.} Since different stakeholders interact with cybersecurity content in distinct ways, we adopt
a persona-based approach~\citep{nvidia/Nemotron-Personas-USA}. We first prompt the model to
propose a set of personas relevant to each seed document (e.g., a security officer, a
software engineer working in JavaScript, a red team operator analyzing a CVE, or a student
learning defensive security), and then simulate a conversation between the selected persona
and a general cybersecurity expert.

\subsubsection{Data Verification and Filtering}\label{sec:verify}
Ensuring high-quality synthetic data is critical, as artifacts introduced during training can
directly shape model behavior, and even a small amount of low-quality data can be
harmful~\citep{Li-etal-2024-superfiltering}. We therefore apply an LLM-based verification
stage to the generated samples using Qwen-30B-A3B. The verifier assigns six sub-scores on a 1--10 scale,
covering instruction adherence, task completion, factuality, format and style alignment,
relevance and focus, and logical consistency and coherence, along with an overall quality
score. Samples with an overall score below $8$ are removed from the final training corpus.

\subsection{General Data}\label{sec:general_mix}
Our synthetic dataset is entirely cybersecurity-centric. To keep the intermediate distribution from becoming a domain-only CPT corpus and to introduce instruction formatting, we augment the training data with general supervised instruction data. We use the publicly available Olmo~3 Dolci SFT dataset~\citep{olmo2025olmo}. A small subset of 50K samples is included during mid-training using a replay strategy~\citep{shi2025continual}, with the rest used during post-training.

\subsection{Training}\label{sec:training}
For MiST-8B, we initialize from the pre-trained \texttt{Qwen3-8B-Base} checkpoint.
For MiST-32B, we initialize from a post-trained Qwen3-32B checkpoint, since a
\texttt{Qwen3-32B-Base} checkpoint was not publicly available at the time of training.
Both models are then adapted using the three-stage MiST pipeline consisting of
cybersecurity mid-training, supervised fine-tuning, and preference tuning via
DPO~\citep{rafailov2023direct}, as shown in \Cref{fig:midtrain_flow}.
During mid-training, we optimize the standard causal language modeling objective, i.e.,
next-token prediction, over the mixed cybersecurity and general-data corpus.
All experiments are conducted on a single node with $8\times$
NVIDIA B200 GPUs.


We run mid-training for two epochs with a maximum sequence length of 16{,}384 tokens and an effective per-device batch size of 16 with sequence packing enabled. We use AdamW~\citep{loshchilov2017decoupled} with a cosine learning rate schedule, with a peak of $5\times10^{-5}$ and a minimum of $1\times10^{-6}$, a warmup ratio of 0.03, weight decay of 0.05, and gradient clipping at 0.2. Training is carried out in bfloat16 precision with FlashAttention-2~\citep{dao2023flashattention} and DeepSpeed~\citep{rajbhandari2020zero}, using ZeRO-2 for the 8B model and ZeRO-3 for the 32B model. The SFT and DPO stages use similar configuration; full hyperparameters and estimated training times are provided in \Cref{tab:train_hparams,tab:train_time}. For the DPO phase we use the general Dolci-Instruct-DPO dataset\footnote{\url{https://huggingface.co/datasets/allenai/Dolci-Instruct-DPO}}\citep{olmo2025olmo}. Because these preference data are not cybersecurity-specific, the DPO stage is intended primarily as a general alignment step rather than as a mechanism for adding new cybersecurity capabilities.

\begin{table*}[t]
\centering
\scriptsize
\setlength{\tabcolsep}{2.6pt}
\renewcommand{\arraystretch}{0.92}
\resizebox{\textwidth}{!}{%
\begin{tabular}{@{}l@{\hspace{0.35em}}cccccccccccccc@{}}
\toprule
 & \multicolumn{2}{c}{\textbf{CTI}} & \multicolumn{3}{c}{\textbf{Secure}} & \multicolumn{4}{c}{\textbf{Athena}} & \multicolumn{2}{c}{\textbf{PrimeVul}} &  &  &  \\[-0.5ex]
\cmidrule(lr){2-3} \cmidrule(lr){4-6} \cmidrule(lr){7-10} \cmidrule(lr){11-12} \noalign{\vskip-1.0ex}
\textbf{Model} & MCQ & RCM & MAET & CWET & KCV & CKT & ATE & RMS & TAA & P-C & P-C CoT & \shortstack{MMLU\\Cyber} & \shortstack{Sec\\Eval} & Mean \\
\midrule
\multicolumn{15}{l}{\textit{Large/Proprietary}} \\[-0.2ex]
GPT-5.4-mini & 73.9 & 73.1 & 91.3 & 92.0 & \underline{88.9} & 85.2 & 53.2 & \underline{29.2} & \textbf{29.8} & 8.4 & 10.1 & 86.0 & 74.7 & 61.2 \\
GPT-5.4-nano & 62.1 & 69.0 & 81.5 & 84.0 & 85.6 & 79.5 & 45.3 & 5.5 & 20.0 & 9.0 & 9.2 & 75.8 & 67.3 & 53.4 \\
Qwen3-235B & 71.9 & 71.4 & 89.6 & 91.5 & 86.2 & 83.6 & 43.5 & 13.1 & 27.6 & 4.5 & 14.6 & \underline{87.0} & 71.3 & 58.1 \\
\midrule
\multicolumn{15}{l}{\textit{Cybersecurity models}} \\[-0.2ex]
DeepHat-7B & 54.7 & 67.2 & 82.7 & 83.1 & 86.2 & 69.2 & 9.8 & 3.1 & 14.8 & \underline{12.0} & 3.9 & 79.8 & 62.4 & 48.4 \\
Foundation-Sec-8B & 58.2 & 67.9 & 84.7 & 83.3 & 80.1 & 76.9 & 37.3 & 18.7 & 26.4 & 0.0 & \underline{17.8} & 78.6 & 65.0 & 53.5 \\
Lily-Cyber-7B & 42.1 & 42.9 & 55.8 & 54.4 & 43.2 & 67.0 & 3.0 & 2.1 & 12.4 & 0.0 & 7.1 & 68.8 & 48.9 & 34.4 \\
Qwen3-8B-Primus & 66.8 & 64.4 & 85.9 & 87.0 & 75.8 & 76.4 & 33.9 & 11.5 & 24.6 & 0.3 & 17.1 & 86.0 & 61.8 & 53.2 \\
RedSage-8B-DPO & 62.0 & 73.2 & 89.6 & 91.2 & 81.1 & 78.8 & 51.6 & 18.8 & 22.8 & 1.6 & 14.8 & 82.4 & 67.4 & 56.5 \\
Primus-8B & 54.7 & 67.2 & 81.9 & 82.6 & 82.0 & 73.7 & 30.6 & 7.3 & 21.4 & 0.3 & 13.4 & 82.2 & 60.9 & 50.6 \\
Primus-70B & 67.7 & 65.1 & 91.4 & \underline{93.7} & 88.1 & 82.1 & 52.7 & 14.8 & 2.2 & 1.0 & 17.6 & \textbf{87.6} & 70.5 & 56.5 \\
CyberPal2.0-20B & 73.8 & 72.8 & 91.4 & 92.5 & 83.6 & 82.7 & 57.0 & 23.3 & 19.6 & 0.8 & 14.1 & 82.2 & 67.1 & 58.5 \\
\midrule
\multicolumn{15}{l}{\textit{Baseline models}} \\[-0.2ex]
Qwen3-8B & 58.2 & 53.5 & 83.8 & 87.5 & 84.5 & 75.2 & 9.4 & 5.5 & 15.0 & 2.0 & 13.0 & 78.0 & 66.7 & 48.6 \\
Qwen3-32B & 68.4 & 65.9 & 89.5 & 91.9 & 88.8 & 80.2 & 26.0 & 6.3 & 15.2 & 4.0 & 16.0 & 84.8 & 72.7 & 54.6 \\
\midrule
\multicolumn{15}{l}{\textbf{Ours}} \\[-0.2ex]
\textbf{MiST-8B-SFT} & 78.9 & 74.1 & 93.7 & 92.7 & 85.4 & 86.0 & \textbf{60.2} & \textbf{29.4} & 23.8 & \textbf{14.0} & \textbf{18.0} & 82.8 & 72.2 & 62.4 \\
\textbf{MiST-8B-DPO} & 79.3 & 73.9 & 94.3 & 93.7 & 87.3 & 86.0 & \underline{57.3} & 25.7 & 24.6 & 10.0 & 15.0 & 83.4 & 71.5 & 61.7 \\
\textbf{MiST-32B-SFT} & \underline{82.2} & \textbf{76.4} & \underline{95.0} & 92.9 & 88.7 & \underline{89.2} & 52.8 & 27.1 & \underline{27.8} & 10.0 & 17.0 & 82.2 & \textbf{76.2} & \underline{62.9} \\
\textbf{MiST-32B-DPO} & \textbf{82.7} & \underline{76.3} & \textbf{95.4} & \textbf{94.1} & \textbf{89.4} & \textbf{89.4} & 55.2 & 28.3 & 26.4 & 8.0 & 15.0 & 84.8 & \underline{76.0} & \textbf{63.2} \\
\bottomrule
\end{tabular}}
\renewcommand{\arraystretch}{1.0}
\caption{Cybersecurity benchmark results with reasoning/thinking modes disabled when available. The Mean column reports the unweighted average across all benchmark columns. \textbf{Bold} indicates best, \underline{underline} indicates second best.}
\label{tab:cyber_results}
\end{table*}

\section{Evaluation}\label{sec:cyber_benchmarks}
We evaluate our model on a broad set of widely used cybersecurity benchmarks to assess its performance across representative security tasks. In addition, we measure performance on general-domain benchmarks to quantify the extent of catastrophic forgetting~\citep{french1999catastrophic,kirkpatrick2017overcoming} resulting from domain-specific training. The cybersecurity suite covers threat-intelligence knowledge, realistic advisory settings, security multiple-choice exams, computer-security knowledge, vulnerability detection, and applied cyber threat intelligence tasks using CTI-Bench~\citep{alam2024ctibench}, SECURE~\citep{bhusal2024secure}, SecEval~\citep{li2023seceval}, MMLU-Cyber~\citep{hendrycks2020measuring}, PrimeVul~\citep{primevul}, and AthenaBench~\citep{athenabench}. Full descriptions of these benchmarks are given in \Cref{app:cyber_benchmark_details}.

\paragraph{Contamination controls.}
Cybersecurity benchmarks frequently derive examples from the same public artifacts used to build domain-training corpora, making both identifier and textual leakage important risks. We therefore apply decontamination at three points in the data lifecycle. First, before synthetic generation, we audit the seed corpus against the evaluation suite and remove records that anchor benchmark examples. In particular, we exclude CVE records appearing in the CTI-Bench root-cause-mapping and vulnerability-severity-prediction splits, and remove GHSA-linked code records whose fixing commit hashes overlap with the PrimeVul validation or test sets. All removals and associated identifiers are logged for reproducibility. Second, because most evaluated tasks are multiple-choice, our synthetic flows deliberately avoid MCQ prompts and A/B/C/D answer structures; instead, they generate free-form rewrites, open-ended QA, conversations, and analytical explanations. Third, we apply an identical 13-gram filter to the mid-training, SFT, and DPO datasets before tokenization. The comparison corpus includes benchmark prompts, reference answers, task instructions, and individual messages from multi-turn examples. Text is lower-cased and stripped of non-alphanumeric characters, and each training sample is serialized across all relevant fields and turns; a sample is removed upon a single matching 13-gram. The resulting corpus contains $27{,}201$ benchmark segments and approximately $467{,}000$ unique 13-grams. Across the combined mid-training and SFT data, the full pipeline removes approximately $3.78\%$ of candidate samples. These controls target exact and lightly modified overlap, but cannot rule out paraphrased or semantically equivalent leakage. Full implementation details, reports, and residual limitations are provided in \Cref{app:decontamination}.

For all benchmarks, we report mean performance over five runs. Evaluations use
non-reasoning inference whenever the model or serving API exposes a separate reasoning or
thinking mode, and scoring uses task-appropriate extraction and metrics. Full benchmark
descriptions, dataset sizes, prompting strategies, inference settings, and model
identifiers are provided in
\Cref{app:eval_configs,app:cyber_benchmark_details,tab:general-benchmarks,tab:cyber-benchmarks,tab:prompting-strategies,tab:model_urls}.

\section{Results}\label{sec:results}
We evaluate MiST models across cybersecurity benchmarks, general capabilities, comparisons to raw continual pre-training, and downstream task adaptation. Overall, the results show that MiST improves cybersecurity performance while preserving general abilities and providing a stronger initialization for subsequent specialization.

\subsection{Cybersecurity Results.}
As shown in \Cref{tab:cyber_results}, MiST substantially improves over the corresponding Qwen3 baselines at both model scales, and these gains are already obtained by the SFT checkpoints. MiST-8B-SFT reaches a mean cybersecurity score of 62.4, compared with 48.6 for Qwen3-8B, an absolute improvement of +13.8 percentage points (+28.4\% relative). MiST-32B-SFT reaches 62.9, compared with 54.6 for Qwen3-32B, an absolute improvement of +8.3 percentage points (+15.2\% relative). The final DPO checkpoints obtain similar scores: 61.7 for MiST-8B-DPO, corresponding to +13.1 percentage points (+27.0\% relative), and 63.2 for MiST-32B-DPO, corresponding to +8.6 percentage points (+15.8\% relative). Thus, the main source of cybersecurity improvement is the mid-training and SFT stages, rather than DPO.

The gains are broad rather than concentrated in a single benchmark. MiST improves
most strongly on CTI-Bench and Athena, indicating better performance on both
cyber threat intelligence questions and applied security analysis tasks.
Improvements on SECURE and PrimeVul are smaller but generally positive,
suggesting that the benefits extend across benchmark families.

After cybersecurity mid-training and supervised fine-tuning, the SFT checkpoints
already account for nearly all of the final cybersecurity performance.
MiST-8B-SFT reaches a mean score of 62.4, close to the DPO checkpoint at 61.7,
while MiST-32B-SFT reaches 62.9 compared with 63.2 after DPO. Since DPO uses
general preference data, these results suggest that it can improve final
instruction-following behavior while largely retaining the cybersecurity
capabilities acquired during earlier training stages.

\subsection{General Benchmarks.}
The full list of benchmarks is provided in \Cref{app:general_bench}, with results reported
in \Cref{tab:general_results_std}. We observe that the MiST models largely retain or improve their performance on general non-cybersecurity benchmarks. We hypothesize that this is due to the inclusion of general instruction data during mid-training, SFT and DPO.


\subsection{Mid-training vs. Raw CPT}
\begin{figure}[h]
    \centering
    \includegraphics[width=0.85\linewidth]{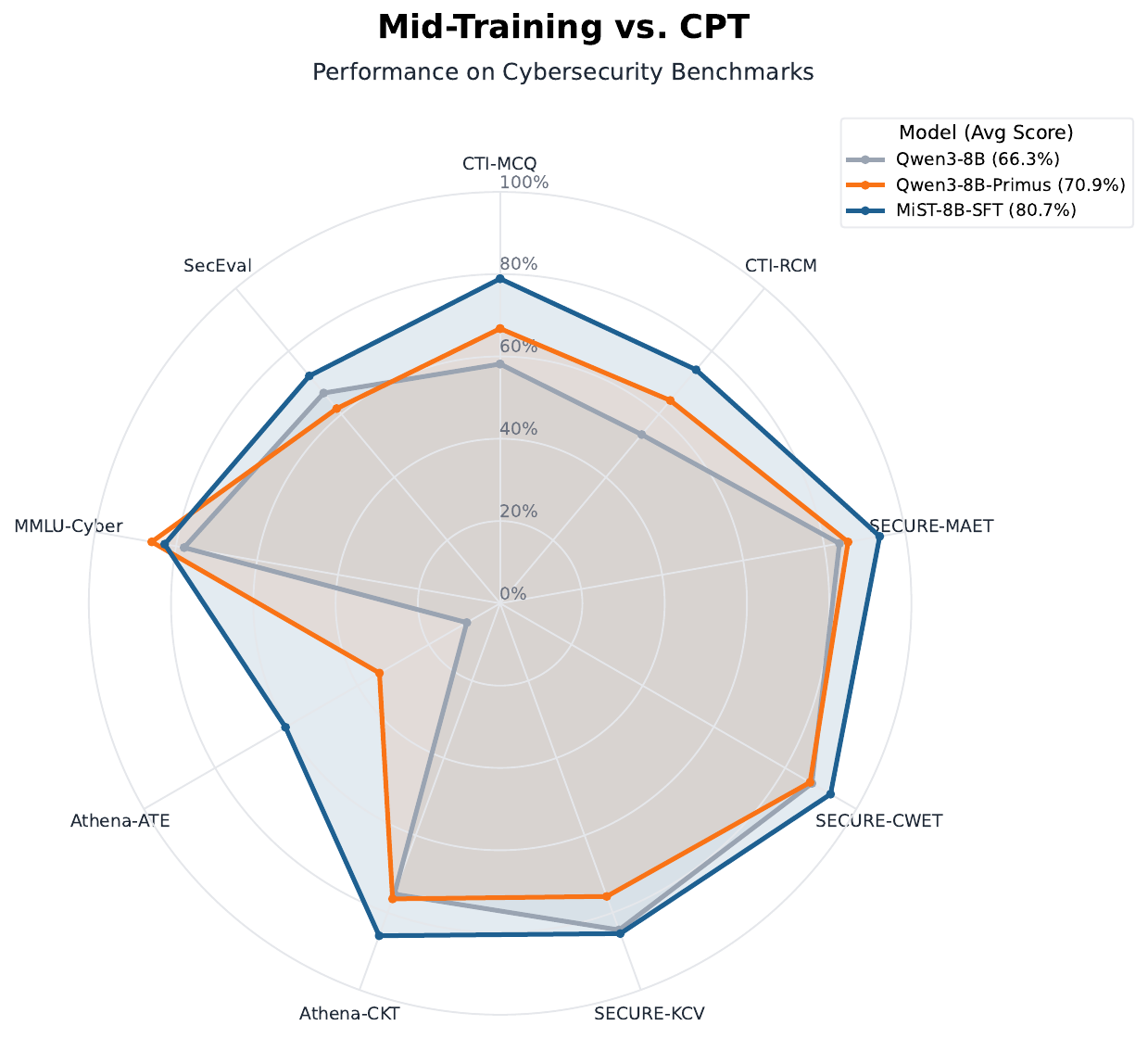}
    \caption{Performance comparison of MiST-8B against the original Qwen3-8B and an 8B raw-CPT baseline.}
    \label{fig:radar}
\end{figure}

\begin{table}[!ht]
\centering
\small
\setlength{\tabcolsep}{2.2pt}
\renewcommand{\arraystretch}{0.94}
\begin{tabular}{l@{\hspace{0.5em}}ccccc}
\toprule
Model & ARC-C & GSM8K & MMLU & IFEval & Mean \\
\midrule
Qwen3-8B & 77.1 & 89.7 & 73.4 & 90.0 & 82.5 \\
MiST-8B-SFT & 86.0 & 88.9 & 74.5 & 88.4 & 84.5 \\
MiST-8B-DPO & 89.8 & 90.0 & 74.9 & 88.5 & 85.8 \\
\midrule
Qwen3-32B & 89.8 & 94.0 & 81.6 & 91.1 & 89.1 \\
MiST-32B-SFT & 94.3 & 93.4 & 81.8 & 90.2 & 89.9 \\
MiST-32B-DPO & 96.2 & 93.5 & 82.1 & 90.9 & 90.7 \\
\bottomrule
\end{tabular}
\caption{Performance of MiST and its Qwen baselines on standard non-cybersecurity benchmarks evaluating reasoning, instruction following, and knowledge. The full list of benchmarks is described in \Cref{app:general_bench}.}
\label{tab:general_results_std}
\end{table}

 To further assess our approach, we compare it to a raw Continual Pre-Training (CPT) baseline. This baseline follows the raw-domain adaptation recipe: continue training on a large cybersecurity text corpus, then apply the same SFT procedure. We construct the raw CPT corpus by augmenting our seed data with PRIMUS~\citep{yu-etal-2025-primus}, yielding $\sim$2.4B tokens, approximately $2\times$ larger than our mid-training corpus (\Cref{tab:data_stats}).

 We train Qwen3-8B-Base on this raw corpus and then apply the same SFT procedure used for MiST-8B, using identical hyperparameters (\Cref{app:hyperparms}). As shown in \Cref{fig:radar}, while raw CPT improves over the original model, MiST achieves superior performance on nearly all benchmarks despite using fewer training tokens. This shows that a compact synthetic mid-training corpus can be more effective than substantially larger raw-domain training for cybersecurity adaptation. Further ablation studies isolating other components of the full methods are shown in \Cref{app:ablations}. 

\subsection{Downstream Task Adaptation}
\label{sec:downstream_adaptation}

We next ask whether cybersecurity mid-training provides a stronger initialization for
downstream task-specific adaptation. We study two settings: reinforcement learning with
verifiable rewards and supervised fine-tuning.

\subsubsection{Task-Specific Reinforcement Learning}

\begin{figure}[!hbt]
    \centering
    \includegraphics[width=1\linewidth]{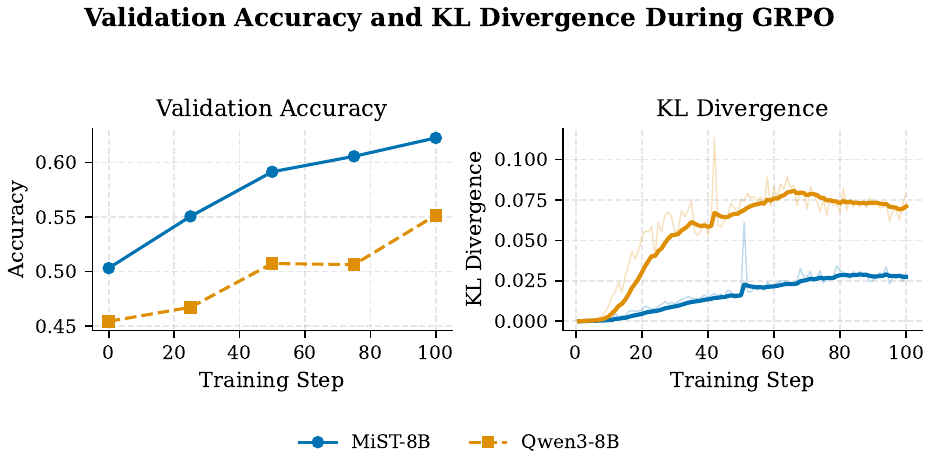}
    \caption{Validation accuracy and tracked KL divergence during GRPO on three
verifiable cybersecurity tasks. Compared with Qwen3-8B, MiST-8B-DPO reaches higher
validation accuracy at every checkpoint while requiring smaller policy movement.}
    \label{fig:rl}
\end{figure}

We first evaluate whether MiST is a better starting point for task-specific reinforcement
learning; full task and reward details are provided in \Cref{app:rl}. We apply
GRPO~\citep{shao2024deepseekmath,mroueh2025grpo} to MiST-8B-DPO and
the original Qwen3-8B using the same RL pipeline on three verifiable cybersecurity tasks:
CVE-to-CWE mapping, CVE-to-CVSS vector prediction, and vulnerable-code-to-CWE mapping.
Because each task can be checked against canonical labels or schemas, rewards are
deterministic and do not require a learned reward model, as in recent cyber threat
intelligence work~\citep{alam2026minerva}.

As shown in \Cref{fig:rl}, RL from the MiST initialization achieves higher validation
accuracy throughout training while maintaining lower KL divergence. This suggests that
cybersecurity mid-training moves the model closer to the downstream task distribution
before RL begins~\citep{tu2025survey,liu2025midtrainingbridges}.
RL can therefore refine cybersecurity behaviors that are already partially present, whereas
the base model must undergo a larger policy shift to reach reward-relevant regions.

This finding is consistent with studies showing that mid-training improves downstream RL
and that RL is most effective near the model's existing competence boundary
~\citep{zhang2025interplay,wang2025octothinker}. It also aligns with analyses of GRPO in
which performance depends on the initial probability of success ~\citep{mroueh2025grpo}. Overall, MiST is not only stronger
before adaptation; it also provides a better RL initialization for specialized
cybersecurity tasks.

\subsubsection{Task-Specific SFT}

We next evaluate whether cybersecurity mid-training also improves supervised downstream
adaptation. We fine-tune Qwen and MiST checkpoints on PrimeVul~\citep{primevul}, a
paired vulnerability-detection benchmark in which each vulnerable function is matched with
a patched counterpart. We then evaluate on the held-out PrimeVul paired test split under
two prompting variants: \textsc{PrimeVul}, which uses a direct YES/NO vulnerability
classification prompt, and \textsc{PrimeVul-CoT}, which asks the model to reason
step-by-step before giving its final verdict. Full training and evaluation details are
provided in \Cref{app:primevul_sft}.


\begin{figure}[!hbt]
  \centering
  \includegraphics[width=0.9\linewidth]{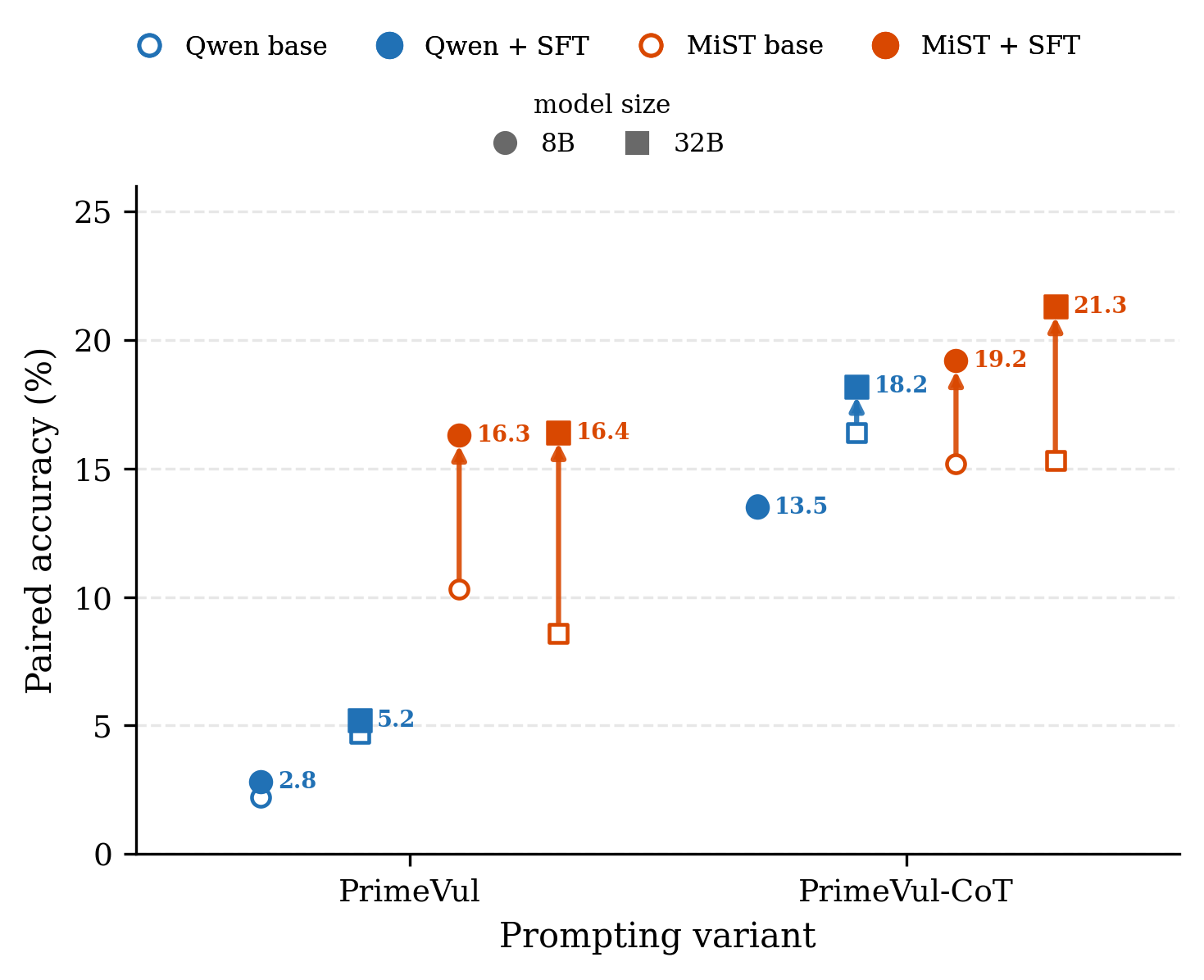}
  \caption{Effect of task-specific SFT on PrimeVul paired accuracy. Each
  vertical pair connects a base checkpoint (hollow marker) to its
  PrimeVul-fine-tuned counterpart (filled marker); arrow length is the gain
  from SFT. Color encodes family (\textcolor[HTML]{2171b5}{\textbf{Qwen}} /
  \textcolor[HTML]{D94801}{\textbf{MiST}}) and marker shape encodes scale
  ($\circ$ = 8B, $\square$ = 32B), shown under direct (\textsc{PrimeVul}) and
  chain-of-thought (\textsc{PrimeVul-CoT}) prompting. MiST arrows are
  consistently longer than Qwen arrows at both scales and under both
  prompting variants, indicating that cybersecurity mid-training provides a
  stronger initialization for downstream task-specific SFT.}
  \label{fig:primevul_sft_slope}
\end{figure}

As shown in \Cref{fig:primevul_sft_slope}, task-specific SFT yields substantially larger gains
when applied to MiST than when applied to Qwen. This holds for both 8B and 32B models and
under both direct and chain-of-thought prompting. These results mirror the RL findings:
    cybersecurity mid-training provides a stronger initialization for downstream adaptation,
especially when the task requires fine-grained security distinctions that are difficult to
learn from task data alone.

\section{Conclusion} 
We introduce MiST, a suite of 8B and 32B cybersecurity-specialized models and propose a data curation and generation recipe for constructing mid-training corpora from information dense data. Our results show that this curated mid-training approach is more effective than continual pre-training on substantially larger raw domain corpora, highlighting the importance of data quality and structure over token volume alone. MiST outperforms existing open cybersecurity-specific models while preserving general capabilities, and matches the performance of similarly sized proprietary models. 

Furthermore, MiST provides a stronger starting point for downstream adaptation. Under both task-specific SFT and reinforcement learning, MiST adapts more effectively than the original Qwen models; in the RL setting, it achieves higher validation accuracy with lower KL movement. Overall, MiST demonstrates that carefully curated mid-training data can produce capable, generalizable, and adaptable open LLMs for cybersecurity.


\section*{Limitations}

This work has several limitations. First, cybersecurity is a broad domain, and our
evaluation focuses primarily on knowledge-intensive security understanding, cyber threat
intelligence, vulnerability reasoning, and benchmark-style analysis. We do not
comprehensively evaluate MiST in operational settings such as log and telemetry analysis,
incident triage, secure code assistance, tool use, or long-horizon agentic workflows.
Accordingly, our results should be interpreted as evidence of improved cybersecurity
knowledge and structured analysis, rather than as a direct measure of deployment readiness
in live security operations.

Second, although we study both task-specific supervised fine-tuning and reinforcement
learning with verifiable rewards, our RL experiments cover only a small set of structured
cybersecurity tasks. We do not explore reasoning-specialized models, broader RL recipes,
tool-use settings, or agentic reinforcement learning. Future work should test whether
cybersecurity mid-training provides similar benefits for more complex multi-step security
tasks.

Third, MiST relies heavily on LLM-generated synthetic data. While synthetic data is widely
used in modern LLM training, it can introduce systematic biases, artifacts, or
distributional distortions~\citep{long-etal-2024-llms}. We mitigate this risk through
expert review and LLM-based filtering, but we do not fully quantify the effects of
generator choice, prompt design, filtering thresholds, or source mixtures. Moreover,
because the same model family is used for generation and verification, the verifier may
share blind spots with the generator.

Fourth, although we apply source-level filtering and $n$-gram decontamination,
contamination cannot be ruled out completely. Exact- and near-exact-match filtering may
miss paraphrased, reformatted, or semantically equivalent benchmark content. This is
especially relevant in cybersecurity, where benchmarks and training sources often refer to
the same public CVEs, CWEs, ATT\&CK techniques, and vulnerability descriptions. Future
evaluations should include additional held-out, time-split, and privately constructed
benchmarks.

\section*{Ethical Considerations}

\paragraph{Data governance.}
MiST is trained using public and third-party cybersecurity resources, together with
synthetic transformations grounded in those resources. Public availability does not
necessarily imply unrestricted permission to redistribute source text or derived examples,
and cybersecurity sources may contain sensitive operational details, personal information,
disputed threat-actor attributions, or information that later becomes outdated. Before
releasing training artifacts, we will document source provenance, collection dates,
applicable licenses or terms of use, transformation procedures, and known restrictions in
a dataset card. We will release only artifacts for which redistribution is permitted,
screen released data for sensitive content, and provide a mechanism for reporting errors
or requesting correction or removal. Because synthetic generation can preserve or amplify
errors and sensitive details from its inputs, we treat generated samples as governed
derivatives of their source material rather than as independent data.

\paragraph{Model governance and dual use.}
MiST is intended for cybersecurity research and defensive analysis and should not be
treated as an autonomous authority for operational security decisions. Improved
vulnerability analysis, threat-intelligence understanding, and security reasoning may also
lower the barrier to offensive or otherwise harmful use. Moreover, the preliminary safety
evaluation in \Cref{app:safety} does not establish safety under adaptive, multi-turn, or
tool-assisted attacks. Model releases will therefore be accompanied by model cards
describing intended and out-of-scope uses, inherited licensing obligations, evaluation
results and limitations, checkpoint versions, and recommended deployment safeguards.
Release decisions will be informed by dual-use red-teaming, and higher-risk artifacts may
be released through staged or access-controlled mechanisms when warranted. Deployers
should apply human oversight, access controls, logging, monitoring, and domain-specific
legal and organizational review, particularly in live security operations.
\section*{Acknowledgments}
We thank Iyar Zaks, Ran Avnimelech, Shay Geller, and Ofek Ophir for their help in evaluating and testing the final model. We also extend our thanks to Tal Fialkow, whose leadership and support were instrumental throughout this work.

\bibliography{custom}

@inproceedings{yu-etal-2025-primus,
    title = "Primus: A Pioneering Collection of Open-Source Datasets for Cybersecurity {LLM} Training",
    author = "Yu, Yao-Ching  and
      Chiang, Tsun-Han  and
      Tsai, Cheng-Wei  and
      Huang, Chien-Ming  and
      Tsao, Wen-Kwang",
    editor = "Christodoulopoulos, Christos  and
      Chakraborty, Tanmoy  and
      Rose, Carolyn  and
      Peng, Violet",
    booktitle = "Proceedings of the 2025 Conference on Empirical Methods in Natural Language Processing",
    month = nov,
    year = "2025",
    address = "Suzhou, China",
    publisher = "Association for Computational Linguistics",
    url = "https://aclanthology.org/2025.emnlp-main.527/",
    doi = "10.18653/v1/2025.emnlp-main.527",
    pages = "10391--10413",
    ISBN = "979-8-89176-332-6",
}

@article{kirkpatrick2017overcoming,
author = {James Kirkpatrick  and Razvan Pascanu  and Neil Rabinowitz  and Joel Veness  and Guillaume Desjardins  and Andrei A. Rusu  and Kieran Milan  and John Quan  and Tiago Ramalho  and Agnieszka Grabska-Barwinska  and Demis Hassabis  and Claudia Clopath  and Dharshan Kumaran  and Raia Hadsell },
title = {Overcoming catastrophic forgetting in neural networks},
journal = {Proceedings of the National Academy of Sciences},
volume = {114},
number = {13},
pages = {3521-3526},
year = {2017},
doi = {10.1073/pnas.1611835114},
URL = {https://www.pnas.org/doi/abs/10.1073/pnas.1611835114},
eprint = {https://www.pnas.org/doi/pdf/10.1073/pnas.1611835114},
}

@article{french1999catastrophic,
  title={Catastrophic forgetting in connectionist networks},
  author={French, Robert M},
  journal={Trends in cognitive sciences},
  volume={3},
  number={4},
  pages={128--135},
  year={1999},
  publisher={Elsevier}
}

@inproceedings{rajbhandari2020zero,
  title={Zero: Memory optimizations toward training trillion parameter models},
  author={Rajbhandari, Samyam and Rasley, Jeff and Ruwase, Olatunji and He, Yuxiong},
  booktitle={SC20: International Conference for High Performance Computing, Networking, Storage and Analysis},
  pages={1--16},
  year={2020},
  organization={IEEE}
}

@article{wang2025octothinker,
  title={Octothinker: Mid-training incentivizes reinforcement learning scaling},
  author={Wang, Zengzhi and Zhou, Fan and Li, Xuefeng and Liu, Pengfei},
  journal={arXiv preprint arXiv:2506.20512},
  year={2025}
}

@misc{abdin2024phi3technicalreporthighly,
      title={Phi-3 Technical Report: A Highly Capable Language Model Locally on Your Phone}, 
      author={Marah Abdin and Jyoti Aneja and Hany Awadalla and Ahmed Awadallah and Ammar Ahmad Awan and Nguyen Bach and Amit Bahree and Arash Bakhtiari and Jianmin Bao and Harkirat Behl and Alon Benhaim and Misha Bilenko and Johan Bjorck and Sébastien Bubeck and Martin Cai and Qin Cai and Vishrav Chaudhary and Dong Chen and Dongdong Chen and Weizhu Chen and Yen-Chun Chen and Yi-Ling Chen and Hao Cheng and Parul Chopra and Xiyang Dai and Matthew Dixon and Ronen Eldan and Victor Fragoso and Jianfeng Gao and Mei Gao and Min Gao and Amit Garg and Allie Del Giorno and Abhishek Goswami and Suriya Gunasekar and Emman Haider and Junheng Hao and Russell J. Hewett and Wenxiang Hu and Jamie Huynh and Dan Iter and Sam Ade Jacobs and Mojan Javaheripi and Xin Jin and Nikos Karampatziakis and Piero Kauffmann and Mahoud Khademi and Dongwoo Kim and Young Jin Kim and Lev Kurilenko and James R. Lee and Yin Tat Lee and Yuanzhi Li and Yunsheng Li and Chen Liang and Lars Liden and Xihui Lin and Zeqi Lin and Ce Liu and Liyuan Liu and Mengchen Liu and Weishung Liu and Xiaodong Liu and Chong Luo and Piyush Madan and Ali Mahmoudzadeh and David Majercak and Matt Mazzola and Caio César Teodoro Mendes and Arindam Mitra and Hardik Modi and Anh Nguyen and Brandon Norick and Barun Patra and Daniel Perez-Becker and Thomas Portet and Reid Pryzant and Heyang Qin and Marko Radmilac and Liliang Ren and Gustavo de Rosa and Corby Rosset and Sambudha Roy and Olatunji Ruwase and Olli Saarikivi and Amin Saied and Adil Salim and Michael Santacroce and Shital Shah and Ning Shang and Hiteshi Sharma and Yelong Shen and Swadheen Shukla and Xia Song and Masahiro Tanaka and Andrea Tupini and Praneetha Vaddamanu and Chunyu Wang and Guanhua Wang and Lijuan Wang and Shuohang Wang and Xin Wang and Yu Wang and Rachel Ward and Wen Wen and Philipp Witte and Haiping Wu and Xiaoxia Wu and Michael Wyatt and Bin Xiao and Can Xu and Jiahang Xu and Weijian Xu and Jilong Xue and Sonali Yadav and Fan Yang and Jianwei Yang and Yifan Yang and Ziyi Yang and Donghan Yu and Lu Yuan and Chenruidong Zhang and Cyril Zhang and Jianwen Zhang and Li Lyna Zhang and Yi Zhang and Yue Zhang and Yunan Zhang and Xiren Zhou},
      year={2024},
      eprint={2404.14219},
      archivePrefix={arXiv},
      primaryClass={cs.CL},
      url={https://arxiv.org/abs/2404.14219}, 
}

@misc{wake2024yi,
  title={Yi-lightning technical report},
  author={Alan Wake and Bei Chen and C. X. Lv and Chao Li and Chengen Huang and Chenglin Cai and Chujie Zheng and Daniel Cooper and Fan Zhou and Feng Hu and Ge Zhang and Guoyin Wang and Heng Ji and Howard Qiu and Jiangcheng Zhu and Jun Tian and Katherine Su and Lihuan Zhang and Liying Li and Ming Song and Mou Li and Peng Liu and Qicheng Hu and Shawn Wang and Shijun Zhou and Shiming Yang and Shiyong Li and Tianhang Zhu and Wen Xie and Wenhao Huang and Xiang He and Xiaobo Chen and Xiaohui Hu and Xiaoyi Ren and Xinyao Niu and Yanpeng Li and Yongke Zhao and Yongzhen Luo and Yuchi Xu and Yuxuan Sha and Zhaodong Yan and Zhiyuan Liu and Zirui Zhang and Zonghong Dai},
      year={2025},
      eprint={2412.01253},
      archivePrefix={arXiv},
      primaryClass={cs.CL},
      url={https://arxiv.org/abs/2412.01253}, 
}

@misc{tu2025survey,
      title={A Survey on LLM Mid-Training}, 
      author={Chengying Tu and Xuemiao Zhang and Rongxiang Weng and Rumei Li and Chen Zhang and Yang Bai and Hongfei Yan and Jingang Wang and Xunliang Cai},
      year={2025},
      eprint={2510.23081},
      archivePrefix={arXiv},
      primaryClass={cs.CL},
      url={https://arxiv.org/abs/2510.23081}, 
}

@article{li2023textbooks,
  title={Textbooks are all you need ii: phi-1.5 technical report},
  author={Li, Yuanzhi and Bubeck, S{\'e}bastien and Eldan, Ronen and Del Giorno, Allie and Gunasekar, Suriya and Lee, Yin Tat},
  journal={arXiv preprint arXiv:2309.05463},
  year={2023}
}

@inproceedings{penedo2024fineweb,
 author = {Penedo, Guilherme and Kydl\'{\i}\v{c}ek, Hynek and allal, Loubna Ben and Lozhkov, Anton and Mitchell, Margaret and Raffel, Colin and Von Werra, Leandro and Wolf, Thomas},
 booktitle = {Advances in Neural Information Processing Systems},
 doi = {10.52202/079017-0970},
 editor = {A. Globerson and L. Mackey and D. Belgrave and A. Fan and U. Paquet and J. Tomczak and C. Zhang},
 pages = {30811--30849},
 publisher = {Curran Associates, Inc.},
 title = {The FineWeb Datasets: Decanting the Web for the Finest Text Data at Scale},
 url = {https://proceedings.neurips.cc/paper_files/paper/2024/file/370df50ccfdf8bde18f8f9c2d9151bda-Paper-Datasets_and_Benchmarks_Track.pdf},
 volume = {37},
 year = {2024}
}

@article{loshchilov2017decoupled,
  title={Decoupled weight decay regularization},
  author={Loshchilov, Ilya and Hutter, Frank},
  journal={arXiv preprint arXiv:1711.05101},
  year={2017}
}

@article{jiang2024survey,
  title={A survey on large language models for code generation},
  author={Jiang, Juyong and Wang, Fan and Shen, Jiasi and Kim, Sungju and Kim, Sunghun},
  journal={arXiv preprint arXiv:2406.00515},
  year={2024}
}

@article{xu2024large,
author = {Xu, Hanxiang and Wang, Shenao and Li, Ningke and Wang, Kailong and Zhao, Yanjie and Chen, Kai and Yu, Ting and Liu, Yang and Wang, Haoyu},
title = {Large Language Models for Cyber Security: A Systematic Literature Review},
year = {2025},
publisher = {Association for Computing Machinery},
address = {New York, NY, USA},
issn = {1049-331X},
url = {https://doi.org/10.1145/3769676},
doi = {10.1145/3769676},
note = {Just Accepted},
journal = {ACM Trans. Softw. Eng. Methodol.},
month = sep
}

@inproceedings{li-etal-2024-superfiltering,
    title = "Superfiltering: Weak-to-Strong Data Filtering for Fast Instruction-Tuning",
    author = "Li, Ming  and
      Zhang, Yong  and
      He, Shwai  and
      Li, Zhitao  and
      Zhao, Hongyu  and
      Wang, Jianzong  and
      Cheng, Ning  and
      Zhou, Tianyi",
    editor = "Ku, Lun-Wei  and
      Martins, Andre  and
      Srikumar, Vivek",
    booktitle = "Proceedings of the 62nd Annual Meeting of the Association for Computational Linguistics (Volume 1: Long Papers)",
    month = aug,
    year = "2024",
    address = "Bangkok, Thailand",
    publisher = "Association for Computational Linguistics",
    url = "https://aclanthology.org/2024.acl-long.769/",
    doi = "10.18653/v1/2024.acl-long.769",
    pages = "14255--14273",
}

@article{zhou2023instruction,
  title={Instruction-following evaluation for large language models},
  author={Zhou, Jeffrey and Lu, Tianjian and Mishra, Swaroop and Brahma, Siddhartha and Basu, Sujoy and Luan, Yi and Zhou, Denny and Hou, Le},
  journal={arXiv preprint arXiv:2311.07911},
  year={2023}
}

@article{hendrycks2020measuring,
  title={Measuring massive multitask language understanding},
  author={Hendrycks, Dan and Burns, Collin and Basart, Steven and Zou, Andy and Mazeika, Mantas and Song, Dawn and Steinhardt, Jacob},
  journal={arXiv preprint arXiv:2009.03300},
  year={2020}
}

@misc{weerawardhena2025llama,
  title={Llama-3.1-FoundationAI-SecurityLLM-8B-Instruct Technical Report}, 
  author={Sajana Weerawardhena and Paul Kassianik and Blaine Nelson and Baturay Saglam and Anu Vellore and Aman Priyanshu and Supriti Vijay and Massimo Aufiero and Arthur Goldblatt and Fraser Burch and Ed Li and Jianliang He and Dhruv Kedia and Kojin Oshiba and Zhouran Yang and Yaron Singer and Amin Karbasi},
  year={2025},
  eprint={2508.01059},
  archivePrefix={arXiv},
  primaryClass={cs.CR},
  url={https://arxiv.org/abs/2508.01059}
}

@inproceedings{brown2020language,
 author = {Brown, Tom and Mann, Benjamin and Ryder, Nick and Subbiah, Melanie and Kaplan, Jared D and Dhariwal, Prafulla and Neelakantan, Arvind and Shyam, Pranav and Sastry, Girish and Askell, Amanda and Agarwal, Sandhini and Herbert-Voss, Ariel and Krueger, Gretchen and Henighan, Tom and Child, Rewon and Ramesh, Aditya and Ziegler, Daniel and Wu, Jeffrey and Winter, Clemens and Hesse, Chris and Chen, Mark and Sigler, Eric and Litwin, Mateusz and Gray, Scott and Chess, Benjamin and Clark, Jack and Berner, Christopher and McCandlish, Sam and Radford, Alec and Sutskever, Ilya and Amodei, Dario},
 booktitle = {Advances in Neural Information Processing Systems},
 editor = {H. Larochelle and M. Ranzato and R. Hadsell and M.F. Balcan and H. Lin},
 pages = {1877--1901},
 publisher = {Curran Associates, Inc.},
 title = {Language Models are Few-Shot Learners},
 url = {https://proceedings.neurips.cc/paper_files/paper/2020/file/1457c0d6bfcb4967418bfb8ac142f64a-Paper.pdf},
 volume = {33},
 year = {2020}
}

@misc{olmo2025olmo,
      title={Olmo 3}, 
      author={Team Olmo and Allyson Ettinger and Amanda Bertsch and Bailey Kuehl and David Graham and David Heineman and Dirk Groeneveld and Faeze Brahman and Finbarr Timbers and Hamish Ivison and Jacob Morrison and Jake Poznanski and Kyle Lo and Luca Soldaini and Matt Jordan and Mayee Chen and Michael Noukhovitch and Nathan Lambert and Pete Walsh and Pradeep Dasigi and Robert Berry and Saumya Malik and Saurabh Shah and Scott Geng and Shane Arora and Shashank Gupta and Taira Anderson and Teng Xiao and Tyler Murray and Tyler Romero and Victoria Graf and Akari Asai and Akshita Bhagia and Alexander Wettig and Alisa Liu and Aman Rangapur and Chloe Anastasiades and Costa Huang and Dustin Schwenk and Harsh Trivedi and Ian Magnusson and Jaron Lochner and Jiacheng Liu and Lester James V. Miranda and Maarten Sap and Malia Morgan and Michael Schmitz and Michal Guerquin and Michael Wilson and Regan Huff and Ronan Le Bras and Rui Xin and Rulin Shao and Sam Skjonsberg and Shannon Zejiang Shen and Shuyue Stella Li and Tucker Wilde and Valentina Pyatkin and Will Merrill and Yapei Chang and Yuling Gu and Zhiyuan Zeng and Ashish Sabharwal and Luke Zettlemoyer and Pang Wei Koh and Ali Farhadi and Noah A. Smith and Hannaneh Hajishirzi},
      year={2025},
      eprint={2512.13961},
      archivePrefix={arXiv},
      primaryClass={cs.CL},
      url={https://arxiv.org/abs/2512.13961}
}

@article{alam2024ctibench,
  title={Ctibench: A benchmark for evaluating llms in cyber threat intelligence},
  author={Alam, Md Tanvirul and Bhusal, Dipkamal and Nguyen, Le and Rastogi, Nidhi},
  journal={Advances in Neural Information Processing Systems},
  volume={37},
  pages={50805--50825},
  year={2024}
}

@inproceedings{bhusal2024secure,
  title={SECURE: Benchmarking Large Language Models for Cybersecurity},
  author={Bhusal, Dipkamal and Alam, Md Tanvirul and Nguyen, Le and Mahara, Ashim and Lightcap, Zachary and Frazier, Rodney and Fieblinger, Romy and Torales, Grace Long and Blakely, Benjamin A and Rastogi, Nidhi},
  booktitle={2024 Annual Computer Security Applications Conference (ACSAC)},
  pages={15--30},
  year={2024},
  organization={IEEE}
}

@article{rafailov2023direct,
  title={Direct preference optimization: Your language model is secretly a reward model},
  author={Rafailov, Rafael and Sharma, Archit and Mitchell, Eric and Manning, Christopher D and Ermon, Stefano and Finn, Chelsea},
  journal={Advances in neural information processing systems},
  volume={36},
  pages={53728--53741},
  year={2023}
}

@article{zhang2025llms,
  title={When llms meet cybersecurity: A systematic literature review},
  author={Zhang, Jie and Bu, Haoyu and Wen, Hui and Liu, Yongji and Fei, Haiqiang and Xi, Rongrong and Li, Lun and Yang, Yun and Zhu, Hongsong and Meng, Dan},
  journal={Cybersecurity},
  volume={8},
  number={1},
  pages={55},
  year={2025},
  publisher={Springer}
}

@article{minaee2024large,
  title={Large language models: A survey},
  author={Minaee, Shervin and Mikolov, Tomas and Nikzad, Narjes and Chenaghlu, Meysam and Socher, Richard and Amatriain, Xavier and Gao, Jianfeng},
  journal={arXiv preprint arXiv:2402.06196},
  year={2024}
}

@article{li2021comprehensive,
  title={A comprehensive review study of cyber-attacks and cyber security; Emerging trends and recent developments},
  author={Li, Yuchong and Liu, Qinghui},
  journal={Energy Reports},
  volume={7},
  pages={8176--8186},
  year={2021},
  publisher={Elsevier}
}

@article{ling2023domain,
author = {Ling, Chen and Zhao, Xujiang and Lu, Jiaying and Deng, Chengyuan and Zheng, Can and Wang, Junxiang and Chowdhury, Tanmoy and Li, Yun and Cui, Hejie and Zhang, Xuchao and Zhao, Tianjiao and Panalkar, Amit and Mehta, Dhagash and Pasquali, Stefano and Cheng, Wei and Wang, Haoyu and Liu, Yanchi and Chen, Zhengzhang and Chen, Haifeng and White, Chris and Gu, Quanquan and Pei, Jian and Yang, Carl and Zhao, Liang},
title = {Domain Specialization as the Key to Make Large Language Models Disruptive: A Comprehensive Survey},
year = {2025},
issue_date = {February 2026},
publisher = {Association for Computing Machinery},
address = {New York, NY, USA},
volume = {58},
number = {3},
issn = {0360-0300},
url = {https://doi.org/10.1145/3764579},
doi = {10.1145/3764579},
journal = {ACM Comput. Surv.},
month = oct,
articleno = {79},
numpages = {39}
}

@misc{wang2023survey,
  title={Survey on Factuality in Large Language Models: Knowledge, Retrieval and Domain-Specificity}, 
  author={Cunxiang Wang and Xiaoze Liu and Yuanhao Yue and Xiangru Tang and Tianhang Zhang and Cheng Jiayang and Yunzhi Yao and Wenyang Gao and Xuming Hu and Zehan Qi and Yidong Wang and Linyi Yang and Jindong Wang and Xing Xie and Zheng Zhang and Yue Zhang},
  year={2023},
  eprint={2310.07521},
  archivePrefix={arXiv},
  primaryClass={cs.CL},
  url={https://arxiv.org/abs/2310.07521}, 
}

@article{wu2024pmc,
  title={PMC-LLaMA: toward building open-source language models for medicine},
  author={Wu, Chaoyi and Lin, Weixiong and Zhang, Xiaoman and Zhang, Ya and Xie, Weidi and Wang, Yanfeng},
  journal={Journal of the American Medical Informatics Association},
  volume={31},
  number={9},
  pages={1833--1843},
  year={2024},
  publisher={Oxford Academic}
}

@misc{bhatia2024fintral,
  title={FinTral: A Family of GPT-4 Level Multimodal Financial Large Language Models}, 
  author={Gagan Bhatia and El Moatez Billah Nagoudi and Hasan Cavusoglu and Muhammad Abdul-Mageed},
  year={2024},
  eprint={2402.10986},
  archivePrefix={arXiv},
  primaryClass={cs.CL},
  url={https://arxiv.org/abs/2402.10986}, 
}

@misc{hui2024qwen2,
 title={Qwen2.5-Coder Technical Report}, 
  author={Binyuan Hui and Jian Yang and Zeyu Cui and Jiaxi Yang and Dayiheng Liu and Lei Zhang and Tianyu Liu and Jiajun Zhang and Bowen Yu and Keming Lu and Kai Dang and Yang Fan and Yichang Zhang and An Yang and Rui Men and Fei Huang and Bo Zheng and Yibo Miao and Shanghaoran Quan and Yunlong Feng and Xingzhang Ren and Xuancheng Ren and Jingren Zhou and Junyang Lin},
  year={2024},
  eprint={2409.12186},
  archivePrefix={arXiv},
  primaryClass={cs.CL},
  url={https://arxiv.org/abs/2409.12186}, 
}

@misc{guo2024deepseek,
  title={DeepSeek-Coder: When the Large Language Model Meets Programming -- The Rise of Code Intelligence}, 
  author={Daya Guo and Qihao Zhu and Dejian Yang and Zhenda Xie and Kai Dong and Wentao Zhang and Guanting Chen and Xiao Bi and Y. Wu and Y. K. Li and Fuli Luo and Yingfei Xiong and Wenfeng Liang},
  year={2024},
  eprint={2401.14196},
  archivePrefix={arXiv},
  primaryClass={cs.SE},
  url={https://arxiv.org/abs/2401.14196},
}

@inproceedings{huang2025opencoder,
    title = "{O}pen{C}oder: The Open Cookbook for Top-Tier Code Large Language Models",
    author = "Huang, Siming  and
      Cheng, Tianhao  and
      Liu, Jason Klein  and
      Xu, Weidi  and
      Hao, Jiaran  and
      Song, Liuyihan  and
      Xu, Yang  and
      Yang, Jian  and
      Liu, Jiaheng  and
      Zhang, Chenchen  and
      Chai, Linzheng  and
      Yuan, Ruifeng  and
      Luo, Xianzhen  and
      Wang, Qiufeng  and
      Fan, YuanTao  and
      Zhu, Qingfu  and
      Zhang, Zhaoxiang  and
      Gao, Yang  and
      Fu, Jie  and
      Liu, Qian  and
      Li, Houyi  and
      Zhang, Ge  and
      Qi, Yuan  and
      Yinghui, Xu  and
      Chu, Wei  and
      Wang, Zili",
    editor = "Che, Wanxiang  and
      Nabende, Joyce  and
      Shutova, Ekaterina  and
      Pilehvar, Mohammad Taher",
    booktitle = "Proceedings of the 63rd Annual Meeting of the Association for Computational Linguistics (Volume 1: Long Papers)",
    month = jul,
    year = "2025",
    address = "Vienna, Austria",
    publisher = "Association for Computational Linguistics",
    url = "https://aclanthology.org/2025.acl-long.1591/",
    doi = "10.18653/v1/2025.acl-long.1591",
    pages = "33167--33193",
    ISBN = "979-8-89176-251-0",
}

@misc{kassianik2025llama,
  title={Llama-3.1-FoundationAI-SecurityLLM-Base-8B Technical Report}, 
  author={Paul Kassianik and Baturay Saglam and Alexander Chen and Blaine Nelson and Anu Vellore and Massimo Aufiero and Fraser Burch and Dhruv Kedia and Avi Zohary and Sajana Weerawardhena and Aman Priyanshu and Adam Swanda and Amy Chang and Hyrum Anderson and Kojin Oshiba and Omar Santos and Yaron Singer and Amin Karbasi},
  year={2025},
  eprint={2504.21039},
  archivePrefix={arXiv},
  primaryClass={cs.CR},
  url={https://arxiv.org/abs/2504.21039}, 
}

@misc{deephatv1,
  title        = {DeepHat-V1-7B},
  author       = {{DeepHat}},
  year         = {2025},
  howpublished = {\url{https://huggingface.co/DeepHat/}},
  note         = {Model card}
}

@misc{lilycybersecurity,
  title        = {Lily-Cybersecurity-7B-v0.2},
  author       = {{SegoLily Labs}},
  year         = {2025},
  howpublished = {\url{https://huggingface.co/segolilylabs/Lily-Cybersecurity-7B-v0.2}},
  note         = {Model card}
}

@misc{li2023seceval,
    title={SecEval: A Comprehensive Benchmark for Evaluating Cybersecurity Knowledge of Foundation Models},
    author={Li, Guancheng and Li, Yifeng and Wang, Guannan and Yang, Haoyu and Yu, Yang},
    publisher = {GitHub},
    howpublished= "https://github.com/XuanwuAI/SecEval",
    year={2023}
}

@inproceedings{zheng2024sglang,
 author = {Zheng, Lianmin and Yin, Liangsheng and Xie, Zhiqiang and Sun, Chuyue and Huang, Jeff and Yu, Cody Hao and Cao, Shiyi and Kozyrakis, Christos and Stoica, Ion and Gonzalez, Joseph E. and Barrett, Clark and Sheng, Ying},
 booktitle = {Advances in Neural Information Processing Systems},
 doi = {10.52202/079017-2000},
 editor = {A. Globerson and L. Mackey and D. Belgrave and A. Fan and U. Paquet and J. Tomczak and C. Zhang},
 pages = {62557--62583},
 publisher = {Curran Associates, Inc.},
 title = {SGLang: Efficient Execution of Structured Language Model Programs},
 url = {https://proceedings.neurips.cc/paper_files/paper/2024/file/724be4472168f31ba1c9ac630f15dec8-Paper-Conference.pdf},
 volume = {37},
 year = {2024} 
}

@inproceedings{ovadia2024fine,
  title={Fine-tuning or retrieval? comparing knowledge injection in llms},
  author={Ovadia, Oded and Brief, Menachem and Mishaeli, Moshik and Elisha, Oren},
  booktitle={Proceedings of the 2024 conference on empirical methods in natural language processing},
  pages={237--250},
  year={2024}
}

@misc{team2025kimi,
      title={Kimi K2: Open Agentic Intelligence}, 
      author={Kimi Team and Yifan Bai and Yiping Bao and Guanduo Chen and Jiahao Chen and Ningxin Chen and Ruijue Chen and Yanru Chen and Yuankun Chen and Yutian Chen and Zhuofu Chen and Jialei Cui and Hao Ding and Mengnan Dong and Angang Du and Chenzhuang Du and Dikang Du and Yulun Du and Yu Fan and Yichen Feng and Kelin Fu and Bofei Gao and Hongcheng Gao and Peizhong Gao and Tong Gao and Xinran Gu and Longyu Guan and Haiqing Guo and Jianhang Guo and Hao Hu and Xiaoru Hao and Tianhong He and Weiran He and Wenyang He and Chao Hong and Yangyang Hu and Zhenxing Hu and Weixiao Huang and Zhiqi Huang and Zihao Huang and Tao Jiang and Zhejun Jiang and Xinyi Jin and Yongsheng Kang and Guokun Lai and Cheng Li and Fang Li and Haoyang Li and Ming Li and Wentao Li and Yanhao Li and Yiwei Li and Zhaowei Li and Zheming Li and Hongzhan Lin and Xiaohan Lin and Zongyu Lin and Chengyin Liu and Chenyu Liu and Hongzhang Liu and Jingyuan Liu and Junqi Liu and Liang Liu and Shaowei Liu and T. Y. Liu and Tianwei Liu and Weizhou Liu and Yangyang Liu and Yibo Liu and Yiping Liu and Yue Liu and Zhengying Liu and Enzhe Lu and Lijun Lu and Shengling Ma and Xinyu Ma and Yingwei Ma and Shaoguang Mao and Jie Mei and Xin Men and Yibo Miao and Siyuan Pan and Yebo Peng and Ruoyu Qin and Bowen Qu and Zeyu Shang and Lidong Shi and Shengyuan Shi and Feifan Song and Jianlin Su and Zhengyuan Su and Xinjie Sun and Flood Sung and Heyi Tang and Jiawen Tao and Qifeng Teng and Chensi Wang and Dinglu Wang and Feng Wang and Haiming Wang and Jianzhou Wang and Jiaxing Wang and Jinhong Wang and Shengjie Wang and Shuyi Wang and Yao Wang and Yejie Wang and Yiqin Wang and Yuxin Wang and Yuzhi Wang and Zhaoji Wang and Zhengtao Wang and Zhexu Wang and Chu Wei and Qianqian Wei and Wenhao Wu and Xingzhe Wu and Yuxin Wu and Chenjun Xiao and Xiaotong Xie and Weimin Xiong and Boyu Xu and Jing Xu and Jinjing Xu and L. H. Xu and Lin Xu and Suting Xu and Weixin Xu and Xinran Xu and Yangchuan Xu and Ziyao Xu and Junjie Yan and Yuzi Yan and Xiaofei Yang and Ying Yang and Zhen Yang and Zhilin Yang and Zonghan Yang and Haotian Yao and Xingcheng Yao and Wenjie Ye and Zhuorui Ye and Bohong Yin and Longhui Yu and Enming Yuan and Hongbang Yuan and Mengjie Yuan and Haobing Zhan and Dehao Zhang and Hao Zhang and Wanlu Zhang and Xiaobin Zhang and Yangkun Zhang and Yizhi Zhang and Yongting Zhang and Yu Zhang and Yutao Zhang and Yutong Zhang and Zheng Zhang and Haotian Zhao and Yikai Zhao and Huabin Zheng and Shaojie Zheng and Jianren Zhou and Xinyu Zhou and Zaida Zhou and Zhen Zhu and Weiyu Zhuang and Xinxing Zu},
      year={2025},
      eprint={2507.20534},
      archivePrefix={arXiv},
      primaryClass={cs.LG},
      url={https://arxiv.org/abs/2507.20534}
}

@software{nvidia/Nemotron-Personas-USA,
  author = {Meyer, Yev and Corneil, Dane},
  title = {{Nemotron-Personas-USA}: Synthetic Personas Aligned to Real-World Distributions
},
  month = {June},
  year = {2025},
  url = {https://huggingface.co/datasets/nvidia/Nemotron-Personas-USA}
}

@article{dao2023flashattention,
  title={Flashattention-2: Faster attention with better parallelism and work partitioning},
  author={Dao, Tri},
  journal={arXiv preprint arXiv:2307.08691},
  year={2023}
}

@misc{cobbe2021training,
  title={Training Verifiers to Solve Math Word Problems}, 
  author={Karl Cobbe and Vineet Kosaraju and Mohammad Bavarian and Mark Chen and Heewoo Jun and Lukasz Kaiser and Matthias Plappert and Jerry Tworek and Jacob Hilton and Reiichiro Nakano and Christopher Hesse and John Schulman},
  year={2021},
  eprint={2110.14168},
  archivePrefix={arXiv},
  primaryClass={cs.LG},
  url={https://arxiv.org/abs/2110.14168}, 
}

@article{allenai:arc,
      author    = {Peter Clark  and Isaac Cowhey and Oren Etzioni and Tushar Khot and
                    Ashish Sabharwal and Carissa Schoenick and Oyvind Tafjord},
      title     = {Think you have Solved Question Answering? Try ARC, the AI2 Reasoning Challenge},
      journal   = {arXiv:1803.05457v1},
      year      = {2018},
}

@article{shi2025continual,
  title={Continual learning of large language models: A comprehensive survey},
  author={Shi, Haizhou and Xu, Zihao and Wang, Hengyi and Qin, Weiyi and Wang, Wenyuan and Wang, Yibin and Wang, Zifeng and Ebrahimi, Sayna and Wang, Hao},
  journal={ACM Computing Surveys},
  volume={58},
  number={5},
  pages={1--42},
  year={2025},
  publisher={ACM New York, NY}
}

@inproceedings{long-etal-2024-llms,
    title = "On {LLM}s-Driven Synthetic Data Generation, Curation, and Evaluation: A Survey",
    author = "Long, Lin  and
      Wang, Rui  and
      Xiao, Ruixuan  and
      Zhao, Junbo  and
      Ding, Xiao  and
      Chen, Gang  and
      Wang, Haobo",
    editor = "Ku, Lun-Wei  and
      Martins, Andre  and
      Srikumar, Vivek",
    booktitle = "Findings of the Association for Computational Linguistics: ACL 2024",
    month = aug,
    year = "2024",
    address = "Bangkok, Thailand",
    publisher = "Association for Computational Linguistics",
    url = "https://aclanthology.org/2024.findings-acl.658/",
    doi = "10.18653/v1/2024.findings-acl.658",
    pages = "11065--11082",
}

@article{zhang2025interplay,
  title={On the interplay of pre-training, mid-training, and rl on reasoning language models},
  author={Zhang, Charlie and Neubig, Graham and Yue, Xiang},
  journal={arXiv preprint arXiv:2512.07783},
  year={2025}
}

@inproceedings{huang2025middle,
    title = "A Middle Path for On-Premises {LLM} Deployment: Preserving Privacy Without Sacrificing Model Confidentiality",
    author = "Huang, Hanbo  and
      Li, Yihan  and
      Jiang, Bowen  and
      Jiang, Bo  and
      Liu, Lin  and
      Liu, Zhuotao  and
      Sun, Ruoyu  and
      Liang, Shiyu",
    editor = "Christodoulopoulos, Christos  and
      Chakraborty, Tanmoy  and
      Rose, Carolyn  and
      Peng, Violet",
    booktitle = "Proceedings of the 2025 Conference on Empirical Methods in Natural Language Processing",
    month = nov,
    year = "2025",
    address = "Suzhou, China",
    publisher = "Association for Computational Linguistics",
    url = "https://aclanthology.org/2025.emnlp-main.420/",
    doi = "10.18653/v1/2025.emnlp-main.420",
    pages = "8321--8359",
    ISBN = "979-8-89176-332-6",
}

@inproceedings{levi2025cyberpal,
author = {Levi, Matan and Allouche, Yair and Ohayon, Daniel and Puzanov, Anton},
title = {CyberPal.AI: empowering LLMs with expert-driven cybersecurity instructions},
year = {2025},
isbn = {978-1-57735-897-8},
publisher = {AAAI Press},
url = {https://doi.org/10.1609/aaai.v39i23.34618},
doi = {10.1609/aaai.v39i23.34618},
booktitle = {Proceedings of the Thirty-Ninth AAAI Conference on Artificial Intelligence and Thirty-Seventh Conference on Innovative Applications of Artificial Intelligence and Fifteenth Symposium on Educational Advances in Artificial Intelligence},
articleno = {2720},
numpages = {11},
series = {AAAI'25/IAAI'25/EAAI'25}
}

@article{levi2025toward,
  title={Toward Cybersecurity-Expert Small Language Models},
  author={Levi, Matan and Ohayon, Daniel and Blobstein, Ariel and Sagi, Ravid and Molloy, Ian and Allouche, Yair},
  journal={arXiv preprint arXiv:2510.14113},
  year={2025}
}

@inproceedings{suryanto2026redsage,
  title={RedSage: A Cybersecurity Generalist {LLM}},
  author={Suryanto, Naufal and Naseer, Muzammal and Li, Pengfei and Wasim, Syed Talal and Yi, Jinhui and Gall, Juergen and Ceravolo, Paolo and Damiani, Ernesto},
  booktitle={The Fourteenth International Conference on Learning Representations},
  year={2026},
  url={https://openreview.net/forum?id=W4FAenIrQ2},
}

@article{shao2024deepseekmath,
  title={DeepSeekMath: Pushing the Limits of Mathematical Reasoning in Open Language Models},
  author={Shao, Zhihong and Wang, Peiyi and Zhu, Qihao and Xu, Runxin and Song, Junxiao and Zhang, Mingchuan and Li, Y.K. and Wu, Y. and Guo, Daya},
  journal={arXiv preprint arXiv:2402.03300},
  year={2024}
}

@misc{mroueh2025grpo,
  author        = {Mroueh, Youssef},
  title         = {{Reinforcement Learning with Verifiable Rewards: GRPO's Effective Loss, Dynamics, and Success Amplification}},
  year          = {2025},
  note          = {arXiv:2503.06639 [cs.LG]},
  eprint        = {2503.06639},
  archivePrefix = {arXiv},
  primaryClass  = {cs.LG},
  doi           = {10.48550/arXiv.2503.06639},
  url           = {https://arxiv.org/abs/2503.06639}
}

@misc{alam2026minerva,
  author        = {Alam, Md Tanvirul and Piplai, Aritran and Cardei, Ionut and Rastogi, Nidhi and Worth, Jr., Peter J.},
  title         = {{Minerva: Reinforcement Learning with Verifiable Rewards for Cyber Threat Intelligence LLMs}},
  year          = {2026},
  note          = {arXiv:2602.00513 [cs.LG]},
  eprint        = {2602.00513},
  archivePrefix = {arXiv},
  primaryClass  = {cs.LG},
  doi           = {10.48550/arXiv.2602.00513},
  url           = {https://arxiv.org/abs/2602.00513}
}

@misc{liu2025midtrainingbridges,
  author        = {Liu, Emmy and Neubig, Graham and Xiong, Chenyan},
  title         = {{Midtraining Bridges Pretraining and Posttraining Distributions}},
  year          = {2025},
  note          = {arXiv:2510.14865 [cs.CL]},
  eprint        = {2510.14865},
  archivePrefix = {arXiv},
  primaryClass  = {cs.CL},
  doi           = {10.48550/arXiv.2510.14865},
  url           = {https://arxiv.org/abs/2510.14865}
}

@misc{primevul,
      title={Vulnerability Detection with Code Language Models: How Far Are We?}, 
      author={Yangruibo Ding and Yanjun Fu and Omniyyah Ibrahim and Chawin Sitawarin and Xinyun Chen and Basel Alomair and David Wagner and Baishakhi Ray and Yizheng Chen},
      year={2024},
      eprint={2403.18624},
      archivePrefix={arXiv},
      primaryClass={cs.SE},
      url={https://arxiv.org/abs/2403.18624}, 
}

@misc{athenabench,
      title={AthenaBench: A Dynamic Benchmark for Evaluating LLMs in Cyber Threat Intelligence}, 
      author={Md Tanvirul Alam and Dipkamal Bhusal and Salman Ahmad and Nidhi Rastogi and Peter Worth},
      year={2025},
      eprint={2511.01144},
      archivePrefix={arXiv},
      primaryClass={cs.CR},
      url={https://arxiv.org/abs/2511.01144}, 
}

@inproceedings{dodge2021documenting,
    title = "Documenting Large Webtext Corpora: A Case Study on the Colossal Clean Crawled Corpus",
    author = "Dodge, Jesse  and
      Sap, Maarten  and
      Marasovi{\'c}, Ana  and
      Agnew, William  and
      Ilharco, Gabriel  and
      Groeneveld, Dirk  and
      Mitchell, Margaret  and
      Gardner, Matt",
    editor = "Moens, Marie-Francine  and
      Huang, Xuanjing  and
      Specia, Lucia  and
      Yih, Scott Wen-tau",
    booktitle = "Proceedings of the 2021 Conference on Empirical Methods in Natural Language Processing",
    month = nov,
    year = "2021",
    address = "Online and Punta Cana, Dominican Republic",
    publisher = "Association for Computational Linguistics",
    url = "https://aclanthology.org/2021.emnlp-main.98/",
    doi = "10.18653/v1/2021.emnlp-main.98",
    pages = "1286--1305",
}

@misc{touvron2023llama2,
      title={Llama 2: Open Foundation and Fine-Tuned Chat Models}, 
      author={Hugo Touvron and Louis Martin and Kevin Stone and Peter Albert and Amjad Almahairi and Yasmine Babaei and Nikolay Bashlykov and Soumya Batra and Prajjwal Bhargava and Shruti Bhosale and Dan Bikel and Lukas Blecher and Cristian Canton Ferrer and Moya Chen and Guillem Cucurull and David Esiobu and Jude Fernandes and Jeremy Fu and Wenyin Fu and Brian Fuller and Cynthia Gao and Vedanuj Goswami and Naman Goyal and Anthony Hartshorn and Saghar Hosseini and Rui Hou and Hakan Inan and Marcin Kardas and Viktor Kerkez and Madian Khabsa and Isabel Kloumann and Artem Korenev and Punit Singh Koura and Marie-Anne Lachaux and Thibaut Lavril and Jenya Lee and Diana Liskovich and Yinghai Lu and Yuning Mao and Xavier Martinet and Todor Mihaylov and Pushkar Mishra and Igor Molybog and Yixin Nie and Andrew Poulton and Jeremy Reizenstein and Rashi Rungta and Kalyan Saladi and Alan Schelten and Ruan Silva and Eric Michael Smith and Ranjan Subramanian and Xiaoqing Ellen Tan and Binh Tang and Ross Taylor and Adina Williams and Jian Xiang Kuan and Puxin Xu and Zheng Yan and Iliyan Zarov and Yuchen Zhang and Angela Fan and Melanie Kambadur and Sharan Narang and Aurelien Rodriguez and Robert Stojnic and Sergey Edunov and Thomas Scialom},
      year={2023},
      eprint={2307.09288},
      archivePrefix={arXiv},
      primaryClass={cs.CL},
      url={https://arxiv.org/abs/2307.09288},
}

@misc{sainz2023nlp,
      title={NLP Evaluation in trouble: On the Need to Measure LLM Data Contamination for each Benchmark}, 
      author={Oscar Sainz and Jon Ander Campos and Iker García-Ferrero and Julen Etxaniz and Oier Lopez de Lacalle and Eneko Agirre},
      year={2023},
      eprint={2310.18018},
      archivePrefix={arXiv},
      primaryClass={cs.CL},
      url={https://arxiv.org/abs/2310.18018}, 
}

@misc{magar2022data,
      title={Data Contamination: From Memorization to Exploitation}, 
      author={Inbal Magar and Roy Schwartz},
      year={2022},
      eprint={2203.08242},
      archivePrefix={arXiv},
      primaryClass={cs.CL},
      url={https://arxiv.org/abs/2203.08242}, 
}

@misc{deng2024investigating,
      title={Investigating Data Contamination in Modern Benchmarks for Large Language Models}, 
      author={Chunyuan Deng and Yilun Zhao and Xiangru Tang and Mark Gerstein and Arman Cohan},
      year={2024},
      eprint={2311.09783},
      archivePrefix={arXiv},
      primaryClass={cs.CL},
      url={https://arxiv.org/abs/2311.09783}, 
}

@misc{jacovi2023stop,
      title={Stop Uploading Test Data in Plain Text: Practical Strategies for Mitigating Data Contamination by Evaluation Benchmarks}, 
      author={Alon Jacovi and Avi Caciularu and Omer Goldman and Yoav Goldberg},
      year={2023},
      eprint={2305.10160},
      archivePrefix={arXiv},
      primaryClass={cs.CL},
      url={https://arxiv.org/abs/2305.10160}, 
}

@misc{ovadia2025knowledge,
      title={Knowledge-Instruct: Effective Continual Pre-training from Limited Data using Instructions}, 
      author={Oded Ovadia and Meni Brief and Rachel Lemberg and Eitam Sheetrit},
      year={2025},
      eprint={2504.05571},
      archivePrefix={arXiv},
      primaryClass={cs.CL},
      url={https://arxiv.org/abs/2504.05571}, 
}

@misc{riddell2024quantifying,
      title={Quantifying Contamination in Evaluating Code Generation Capabilities of Language Models}, 
      author={Martin Riddell and Ansong Ni and Arman Cohan},
      year={2024},
      eprint={2403.04811},
      archivePrefix={arXiv},
      primaryClass={cs.SE},
      url={https://arxiv.org/abs/2403.04811}, 
}

@misc{yang2024rethinking,
      title={Rethinking Benchmark and Contamination for Language Models with Rephrased Samples}, 
      author={Shuo Yang and Wei-Lin Chiang and Lianmin Zheng and Joseph E. Gonzalez and Ion Stoica},
      year={2023},
      eprint={2311.04850},
      archivePrefix={arXiv},
      primaryClass={cs.CL},
      url={https://arxiv.org/abs/2311.04850}, 
}

@misc{oren2024proving,
      title={Proving Test Set Contamination in Black Box Language Models}, 
      author={Yonatan Oren and Nicole Meister and Niladri Chatterji and Faisal Ladhak and Tatsunori B. Hashimoto},
      year={2023},
      eprint={2310.17623},
      archivePrefix={arXiv},
      primaryClass={cs.CL},
      url={https://arxiv.org/abs/2310.17623}, 
}

@inproceedings{benzaken2022bitfit,
    title = "{B}it{F}it: Simple Parameter-efficient Fine-tuning for Transformer-based Masked Language-models",
    author = "Ben Zaken, Elad  and
      Goldberg, Yoav  and
      Ravfogel, Shauli",
    editor = "Muresan, Smaranda  and
      Nakov, Preslav  and
      Villavicencio, Aline",
    booktitle = "Proceedings of the 60th Annual Meeting of the Association for Computational Linguistics (Volume 2: Short Papers)",
    month = may,
    year = "2022",
    address = "Dublin, Ireland",
    publisher = "Association for Computational Linguistics",
    url = "https://aclanthology.org/2022.acl-short.1/",
    doi = "10.18653/v1/2022.acl-short.1",
    pages = "1--9",
}

@misc{salsa2025,
      title={SALSA: Single-pass Autoregressive LLM Structured Classification}, 
      author={Ruslan Berdichevsky and Shai Nahum-Gefen and Elad Ben Zaken},
      year={2025},
      eprint={2510.22691},
      archivePrefix={arXiv},
      primaryClass={cs.CL},
      url={https://arxiv.org/abs/2510.22691}, 
}

@article{wan2024cyberseceval3,
  title        = {{CyberSecEval 3}: Advancing the Evaluation of Cybersecurity Risks and Capabilities in Large Language Models},
  author       = {Wan, Shengye and Nikolaidis, Cyrus and Song, Daniel and Molnar, David and Crnkovich, James and Grace, Jayson and Bhatt, Manish and Chennabasappa, Sahana and Whitman, Spencer and Ding, Stephanie and Ionescu, Vlad and Li, Yue and Saxe, Joshua},
  journal      = {arXiv preprint arXiv:2408.01605},
  year         = {2024}
}

@article{liu2025purpcode,
  title        = {{PurpCode}: Reasoning for Safer Code Generation},
  author       = {Liu, Jiawei and Diwan, Nirav and Wang, Zhe and Zhai, Haoyu and Zhou, Xiaona and Nguyen, Kiet A. and Yu, Tianjiao and Wahed, Muntasir and Deng, Yinlin and Benkraouda, Hadjer and Wei, Yuxiang and Zhang, Lingming and Lourentzou, Ismini and Wang, Gang},
  journal      = {arXiv preprint arXiv:2507.19060},
  year         = {2025}
}

\newpage
\appendix

\section{Estimated Training Time}
In \Cref{tab:train_time}, we report the estimated wall-clock training time on a single node
with $8\times$B200 NVIDIA GPUs. The full pipeline requires 7.1 hours (56.8 GPU-hours) for
MiST-8B and 42.7 hours (341.3 GPU-hours) for MiST-32B, with supervised fine-tuning
accounting for the largest share of compute.

\begin{table}[!htb]
    \centering
    \small
    \setlength{\tabcolsep}{4pt}
    \renewcommand{\arraystretch}{1.05}
    \begin{tabular}{llrr}
        \toprule
        \textbf{Model} & \textbf{Stage} & \textbf{Time} & \textbf{Hours/GPU-hours} \\
        \midrule
        \multirow{4}{*}{8B}
                       & Mid-training   & 2h 10m        & 2.17 / 17.33             \\
                       & SFT            & 3h 20m        & 3.33 / 26.67             \\
                       & DPO            & 1h 36m        & 1.60 / 12.80             \\
                       & {Total}        & {7h 06m}      & {7.10} / {56.80}          \\
        \midrule
        \multirow{4}{*}{32B}
                       & Mid-training   & 7h 46m        & 7.77  / 62.13            \\
                       & SFT            & 1d 1h 4m      & 25.07 / 200.53           \\
                       & DPO            & 9h 50m        & 9.83  / 78.67            \\
                       & {Total}        & {1d 18h 40m}  & {42.67} / {341.33}       \\
        \bottomrule
    \end{tabular}
    \caption{Estimated wall-clock training time.}
    \label{tab:train_time}
\end{table}

\section{Training Hyperparameters}\label{app:hyperparms}
The full set of hyperparameters is reported in \Cref{tab:train_hparams}. Unless otherwise
noted, all training stages share the same core configuration: trained for 2 epochs, using
the AdamW optimizer with a cosine learning-rate schedule, bfloat16 precision,
FlashAttention-2, and gradient clipping with a threshold of 0.2. Preference optimization
additionally uses a DPO parameter $\beta = 0.1$ and early stopping after 200 steps. The
only difference between mid-training the 8B and 32B models is the use of DeepSpeed ZeRO-3
for the 32B model to reduce memory usage, with all other settings kept identical.

\begin{table*}[!htb]
    \centering
    \small
    \setlength{\tabcolsep}{6pt}
    \renewcommand{\arraystretch}{1.08}
    \begin{tabular}{lccccc}
        \toprule
        Hyperparameter              & Mid-training       & SFT-8B             & SFT-32B            & DPO-8B             & DPO-32B            \\
        \midrule
        Max sequence length         & 16,384             & 8,192              & 8,192              & 8,192              & 8,192              \\
        Per-device batch size       & 4                  & 16                 & 12                 & 16                 & 1                  \\
        Gradient accumulation steps & 4                  & 2                  & 3                  & 4                  & 64                 \\
        Global batch (tokens)       & $\approx$ 2M       & $\approx$ 2M       & $\approx$ 2.36M    & $\approx$ 4.2M     & $\approx$ 4.2M     \\
        Learning rate (peak)        & $5 \times 10^{-5}$ & $3 \times 10^{-5}$ & $3 \times 10^{-5}$ & $3 \times 10^{-7}$ & $1 \times 10^{-7}$ \\
        Learning rate (min)         & $1 \times 10^{-6}$ & $5 \times 10^{-7}$ & $5 \times 10^{-7}$ & 0                  & $0$                \\
        Warmup ratio                & 0.03               & 0.03               & 0.03               & 0.1                & 0.1                \\
        Weight decay                & 0.05               & 0.05               & 0.05               & 0.1                & 0.1                \\
        DeepSpeed stage             & ZeRO-2/3           & ZeRO-2             & ZeRO-3             & ZeRO-2 (offload)   & ZeRO-2 (offload)   \\
        Packing                     & enabled            & enabled            & enabled            & disabled           & disabled           \\
        \bottomrule
    \end{tabular}
    \caption{Stage-specific hyperparameters for mid-training, supervised fine-tuning (SFT), and preference optimization (DPO). Constant settings shared across all stages are described in the text.}
    \label{tab:train_hparams}
\end{table*}

\section{General Benchmarks Details}\label{app:general_bench}
In addition to the cybersecurity benchmarks described in \Cref{sec:cyber_benchmarks}, we
evaluate our model on general-purpose benchmarks that measure broad LLM capabilities. This
evaluation allows us to quantify potential degradation in general performance resulting
from specialization to a narrow domain, which is a key objective of a successful
mid-training run~\citep{tu2025survey}.

\textbf{IFEval}~\citep{zhou2023instruction} evaluates instruction-following behavior in LLMs by measuring adherence to explicit constraints specified in prompts, such as output format, length, stylistic requirements, or prohibited content. Rather than task accuracy, IFEval focuses on precise compliance, making it well suited for assessing alignment in instruction-tuned models.

\textbf{MMLU}~\citep{hendrycks2020measuring} is a large-scale, knowledge-intensive multiple-choice benchmark covering 57 subjects across the sciences, humanities, social sciences, and professional domains. We use MMLU as a coarse measure of general knowledge and reasoning breadth to assess what is lost when focusing on cybersecurity-specific training.


\textbf{ARC-Challenge (ARC-C)}~\citep{allenai:arc} is the Challenge subset of the AI2 Reasoning Challenge, consisting of grade-school science multiple-choice questions filtered to be difficult for simple retrieval and word co-occurrence baselines.

\textbf{GSM8K}~\citep{cobbe2021training} is a dataset of 8.5K human-written grade-school math word problems paired with natural-language solutions, with problems typically requiring multiple steps of arithmetic reasoning. The benchmark targets robustness to linguistic variation and diverse problem structure, rather than memorization of fixed patterns.

\section{Cybersecurity Benchmark Details}
\label{app:cyber_benchmark_details}

\textbf{CTI-Bench}~\citep{alam2024ctibench} evaluates LLM performance on core cyber threat intelligence (CTI) tasks using real-world threat intelligence data. We report results on the multiple-choice question (MCQ) and root cause mapping (RCM) subsets. MCQ measures factual knowledge and fine-grained understanding of CTI concepts, while RCM evaluates the model's ability to identify underlying vulnerability causes by linking CVE records and bug reports to corresponding CWE entries. For the MCQ subset, we additionally report results broken down by its three largest subjects: attack techniques (ATT), software weaknesses (CWE), and attack patterns (CAP).

\textbf{SECURE}~\citep{bhusal2024secure} is a cybersecurity benchmark designed to evaluate LLM performance in realistic advisory settings for critical infrastructure environments. We use three subsets: MAET, CWET, and KCV. MAET and CWET use multiple-choice questions to assess cybersecurity knowledge, while KCV evaluates knowledge understanding through boolean statements grounded in CVE descriptions.

\textbf{SecEval}~\citep{li2023seceval} is a benchmark with over 2,000 multiple-choice questions spanning nine cybersecurity domains, including system, application, and network security. The questions are generated by prompting GPT-4 with authoritative sources such as textbooks and official documentation.

\textbf{MMLU Computer Security}~\citep{hendrycks2020measuring} We additionally report results on the computer security subset of MMLU (see \Cref{app:general_bench}), which consists of 100 multiple-choice questions related to cybersecurity.

\textbf{PrimeVul}~\citep{primevul} is a vulnerability detection benchmark built from
real-world C/C++ functions with high-quality labels, deduplication, and chronological
splits. Following PrimeVul's recommended protocol, we report paired accuracy: a
vulnerable/patched pair is counted as correct only when both functions are classified
correctly. This metric better captures whether a model can distinguish a vulnerability
from its corresponding patch, rather than merely predicting the marginal likelihood of
vulnerability. In \Cref{tab:cyber_results}, P-C denotes this paired classification
accuracy, and P-C CoT denotes the same metric under the benchmark's chain-of-thought
prompting setting.


\textbf{AthenaBench}~\citep{athenabench} is a dynamic cyber threat intelligence benchmark that extends prior CTI evaluation with updated data construction, deduplication, refined metrics, and additional applied-reasoning tasks. We evaluate four AthenaBench tasks: CTI knowledge testing (CKT), attack technique extraction (ATE), risk mitigation strategy selection (RMS), and threat actor attribution (TAA).

\section{Evaluation Protocol}
\label{app:eval_configs}

We evaluate our models on 13 cybersecurity tasks and 4 general-purpose benchmarks.

\begin{table*}[!htb]
    \centering
    \begin{tabular}{llrl}
        \toprule
        \textbf{Benchmark} & \textbf{Source}                                                                   & \textbf{Size} & \textbf{Task Type}                    \\
        \midrule
        MMLU (5-shot)      & \href{https://huggingface.co/datasets/cais/mmlu}{\texttt{cais/mmlu}}              & 14,042        & Multiple-choice QA (57 subjects)      \\
        ARC-Challenge      & \href{https://huggingface.co/datasets/allenai/ai2_arc}{\texttt{allenai/ai2\_arc}} & 2,590         & Science reasoning MCQ               \\
        GSM8K (3-shot)     & \href{https://huggingface.co/datasets/openai/gsm8k}{\texttt{openai/gsm8k}}        & 1,319         & Mathematical reasoning                \\
        IFEval             & \href{https://huggingface.co/datasets/google/IFEval}{\texttt{google/IFEval}}      & 541           & Instruction following                 \\
        \bottomrule
    \end{tabular}
    \caption{General capability benchmarks used for evaluation.}
    \label{tab:general-benchmarks}
\end{table*}

\begin{table*}[!htb]
    \centering
    \small
    \begin{tabular}{llrl}
        \toprule
        \textbf{Benchmark} & \textbf{Source}                                                                           & \textbf{Size} & \textbf{Task Type}                         \\
        \midrule
        MMLU-Cyber         & \href{https://huggingface.co/datasets/cais/mmlu}{\texttt{cais/mmlu}} (computer\_security) & 100           & Security knowledge MCQ                     \\
        CTI-MCQ            & \href{https://huggingface.co/datasets/AI4Sec/cti-bench}{\texttt{AI4Sec/cti-bench}}        & 2,500         & Threat intelligence MCQ                    \\
        CTI-RCM            & \href{https://huggingface.co/datasets/AI4Sec/cti-bench}{\texttt{AI4Sec/cti-bench}}        & 1,000         & Root cause mapping                         \\
        SecEval            & \href{https://huggingface.co/datasets/XuanwuAI/SecEval}{\texttt{XuanwuAI/SecEval}}        & 2,189         & Multi-select security MCQ                  \\
        SECURE-MAET        & SECURE benchmark                                                                          & 1,072         & MITRE ATT\&CK evaluation MCQ               \\
        SECURE-CWET        & SECURE benchmark                                                                          & 964           & CWE evaluation MCQ                         \\
        SECURE-KCV         & SECURE benchmark                                                                          & 466           & CVE verification true/false                \\
        PrimeVul-Base      & PrimeVul~\citep{primevul}                                                                 & 870            & Vulnerable-code binary classification      \\
        PrimeVul-CoT       & PrimeVul~\citep{primevul}                                                                 & 870            & Vulnerable-code classification with CoT    \\
        AthenaBench-CKT    & AthenaBench~\citep{athenabench}                                                          & 3,000            & CTI knowledge testing                      \\
        AthenaBench-ATE    & AthenaBench~\citep{athenabench}                                                          & 500            & Attack technique extraction                \\
        AthenaBench-RMS    & AthenaBench~\citep{athenabench}                                                          & 500            & Risk mitigation strategy selection         \\
        AthenaBench-TAA    & AthenaBench~\citep{athenabench}                                                          & 100            & Threat actor attribution                   \\
        \bottomrule
    \end{tabular}
    \caption{Cybersecurity benchmarks and task subsets used for evaluation.}
    \label{tab:cyber-benchmarks}
\end{table*}

\paragraph{Reasoning Mode.}
All reported evaluations are conducted with reasoning disabled whenever the model or serving
API exposes a separate reasoning or thinking mode. This applies to our Qwen-derived MiST
models, the original Qwen baselines, and all external comparison models for which such a
control is available. Thus, the reported results measure standard non-reasoning inference
rather than explicit test-time reasoning. Prompting strategies that contain phrases such as
``think step by step'' follow the benchmark's prescribed prompt format, but no model is
allowed to use a separate reasoning mode or hidden reasoning budget.

\paragraph{Inference Configuration.}
We serve models using SGLang~\citep{zheng2024sglang} with data parallelism (DP$=8$) on our
$8\times$B200 node, a context length of 16{,}384 tokens, and a maximum generation length of
    4{,}096 tokens. We use temperature $=0.3$, top-$p=0.95$, top-$k=20$, and repetition penalty
$=1.1$ for all evaluations, except for OpenAI GPT-5 models where these controls are not
fully supported. To ensure a fair comparison, all models are evaluated in a non-reasoning
setting: reasoning or thinking modes are disabled for GPT, Qwen, MiST, and other evaluated
models whenever such controls are available. Each benchmark is run $n=5$ times with
different random seeds (0, 1, 2, 3, 4), and we report mean.

\paragraph{Answer Extraction and Scoring.}
We use rule-based answer extraction with regular expressions tailored to each benchmark’s
expected output format. For multiple-choice tasks, we extract the first valid answer choice
(A--D) from the model response and score with accuracy. For mathematical reasoning (GSM8K),
we extract the final numeric answer using the benchmark’s required output pattern and
report exact match. For CWE mapping (CTI-RCM), we extract CWE identifiers via pattern
matching and report exact match. For IFEval, we use the official metric in the loose
setting. All remaining benchmarks are scored with standard accuracy.
This generation-and-extraction protocol matches the benchmark prompting setup. For structured tasks, however, logit-based alternatives such as SALSA~\citep{salsa2025} can classify in a single forward pass by mapping labels to output tokens and scoring their decoder logits. We leave such interfaces for future evaluation.

\paragraph{Aggregation.}
Suite-level scores are computed as the unweighted mean across benchmarks. CTI-MCQ
sub-scores for attack techniques (ATT), software weaknesses (CWE), and attack patterns
(CAP) are reported separately but excluded from suite averages.

\subsection{Prompting Strategies}

\begin{table*}[!htb]
    \centering
    \small
    \begin{tabularx}{\textwidth}{
        >{\raggedright\arraybackslash}p{0.24\textwidth}
        >{\raggedright\arraybackslash}X
    }
        \toprule
        \textbf{Benchmark} & \textbf{Strategy} \\
        \midrule
        MMLU & 5-shot with dev set examples \\
        MMLU-Cyber & 5-shot with dev set examples (Computer Security subset) \\
        GSM8K & 3-shot with solution/answer format \\
        ARC-C & Zero-shot with step-by-step reasoning instruction \\
        IFEval & Direct prompts from dataset \\
        CTI-MCQ (incl.\ subsets) & Zero-shot with thinking instruction suffix \\
        CTI-RCM & Direct prompts from dataset \\
        SecEval & 1-shot example with multi-select instruction \\
        SECURE benchmarks & Direct prompts from dataset \\
        PrimeVul & Zero-shot YES/NO classification with security-expert system prompt \\
        PrimeVul-CoT & Step-by-step reasoning then \texttt{<answer>YES/NO</answer>} verdict \\
        AthenaBench (all tasks) & Direct prompts from dataset (CKT, ATE, RCM, RMS, VSP, TAA) \\
        \bottomrule
    \end{tabularx}
    \caption{Prompting strategies used for each benchmark.}
    \label{tab:prompting-strategies}
\end{table*}

For all benchmarks, we use the official prompts when available. When no official prompt is
provided or multiple configurations exist, we specify the exact prompt used. Below we list
the prompts for benchmarks with custom instructions.

\paragraph{MMLU System Instruction.}
\begin{quote}
    \small
    You are a large language model trained to answer standardized multiple-choice questions covering a wide range of subjects, including mathematics, science, humanities, and social sciences. For each question, analyze the problem carefully and select the single best answer from the given options (A, B, C, or D). Output only the letter corresponding to the correct answer. Do not include explanations or any additional text.
\end{quote}

\paragraph{GSM8K Instruction.}
\begin{quote}
    \small
    Solve the provided math problem. Note: The final answer must be the last word in the response. Only plain numbers are allowed (no currency symbols, percentage signs, etc.)
\end{quote}

\paragraph{ARC-Challenge Instruction.}
\begin{quote}
    \small
    The following is a multiple choice question. Provide your step-by-step reasoning, then give your answer in the format `Answer: X' where X is the answer choice label.
\end{quote}


\paragraph{CTI-MCQ.}
\begin{quote}
    \small
    You are given a multiple-choice question (MCQ) from a Cyber Threat Intelligence (CTI) knowledge benchmark dataset. Your task is to choose the best option among the four provided. Return your answer as a single uppercase letter: A, B, C, or D.

    \{question\}

    \textbf{Important:} The last line of your answer should contain only the single letter corresponding to the best option, with no additional text.
\end{quote}

\paragraph{SecEval System Instruction.}
\begin{quote}
    \small
    Below are multiple-choice questions concerning cybersecurity. Please select the correct answers and respond with the letters ABCD only.
\end{quote}

\section{Ablation Studies}\label{app:ablations}
 
We run three ablations at the 8B scale to understand which parts of the MiST recipe drive the cybersecurity gains. In the first two, we hold the SFT stage fixed so that differences between models reflect only the mid-training corpus; the third varies only the initialization checkpoint. All models are evaluated with the protocol of \Cref{sec:cyber_benchmarks}.
 
\paragraph{Effect of mid-training.}
A natural question is whether the improvements come from mid-training itself or from the SFT mixture alone. To answer it, we apply the identical SFT procedure directly to Qwen3-8B-Base, skipping mid-training, and compare the result against MiST-8B-SFT. As shown in \Cref{tab:ablation_midtraining}, mid-training raises the mean cybersecurity score from 57.6 to 62.4 (+4.8 points). The gains are largest on applied analysis tasks such as ATE (+15.7), PrimeVul paired classification (+11.1), and KCV (+8.8), with a small drop only on MMLU-Cyber ($-$1.5). Together with the raw-CPT comparison in \Cref{sec:results}, these results indicate that the cybersecurity gains come from the mid-training stage itself rather than from the SFT data alone.
 
\begin{table*}[t]
\centering
\scriptsize
\setlength{\tabcolsep}{2.6pt}
\renewcommand{\arraystretch}{0.92}
\begin{tabular}{@{}l@{\hspace{0.35em}}cccccccccccccc@{}}
\toprule
 & \multicolumn{2}{c}{\textbf{CTI}} & \multicolumn{3}{c}{\textbf{Secure}} & \multicolumn{4}{c}{\textbf{Athena}} & \multicolumn{2}{c}{\textbf{PrimeVul}} &  &  &  \\[-0.5ex]
\cmidrule(lr){2-3} \cmidrule(lr){4-6} \cmidrule(lr){7-10} \cmidrule(lr){11-12} \noalign{\vskip-1.0ex}
\textbf{Model} & MCQ & RCM & MAET & CWET & KCV & CKT & ATE & RMS & TAA & P-C & P-C CoT & \shortstack{MMLU\\Cyber} & \shortstack{Sec\\Eval} & Mean \\
\midrule
MiST-8B-SFT (no mid-training) & 74.1 & 67.0 & 92.7 & 90.5 & 76.6 & 83.0 & 44.5 & 27.3 & 22.0 & 2.9 & 14.3 & \textbf{84.3} & 69.0 & 57.6 \\
MiST-8B-SFT (with mid-training) & \textbf{78.9} & \textbf{74.1} & \textbf{93.7} & \textbf{92.7} & \textbf{85.4} & \textbf{86.0} & \textbf{60.2} & \textbf{29.4} & \textbf{23.8} & \textbf{14.0} & \textbf{18.0} & 82.8 & \textbf{72.2} & \textbf{62.4} \\
\midrule
$\Delta$ (mid $-$ no mid) & +4.8 & +7.1 & +1.0 & +2.3 & +8.8 & +3.0 & +15.7 & +2.0 & +1.8 & +11.1 & +3.7 & $-$1.5 & +3.2 & +4.8 \\
\bottomrule
\end{tabular}
\renewcommand{\arraystretch}{1.0}
\caption{Isolating the effect of mid-training. Both models share the identical SFT stage and differ only in whether cybersecurity mid-training is applied first. \textbf{Bold} indicates the better score per column.}
\label{tab:ablation_midtraining}
\end{table*}
 
\paragraph{Synthetic flow ablations.}
We next ask which synthetic flows are responsible for the gains. We ablate the four document-level transformation flows of \Cref{sec:synth}: paraphrasing, educational transformation, QA, and cyber metrics and terminology analysis (\emph{security analysis} below).
We report two variants: \emph{single-flow}, where the mid-training corpus contains a single flow, and \emph{leave-one-out}, where one flow is removed and the rest are kept. Results are shown in \Cref{tab:ablation_flows}.
 
No single flow recovers the full recipe: the best single-flow variant (paraphrasing, 60.3) remains 2.1 points below the full corpus. At the same time, no flow is redundant, as removing any one of them lowers the mean, by 1.4 to 2.8 points. The flows also peak on different benchmarks (e.g., paraphrasing on KCV and educational transformation on RMS), suggesting they are complementary rather than interchangeable. The gains thus come from the composition of the corpus rather than from any single transformation.
 
\begin{table*}[!tp]
\centering
\scriptsize
\setlength{\tabcolsep}{2.6pt}
\renewcommand{\arraystretch}{0.92}
\begin{tabular}{@{}l@{\hspace{0.35em}}cccccccccccccc@{}}
\toprule
 & \multicolumn{2}{c}{\textbf{CTI}} & \multicolumn{3}{c}{\textbf{Secure}} & \multicolumn{4}{c}{\textbf{Athena}} & \multicolumn{2}{c}{\textbf{PrimeVul}} &  &  &  \\[-0.5ex]
\cmidrule(lr){2-3} \cmidrule(lr){4-6} \cmidrule(lr){7-10} \cmidrule(lr){11-12} \noalign{\vskip-1.0ex}
\textbf{Mid-training corpus} & MCQ & RCM & MAET & CWET & KCV & CKT & ATE & RMS & TAA & P-C & P-C CoT & \shortstack{MMLU\\Cyber} & \shortstack{Sec\\Eval} & Mean \\
\midrule
All flows (full recipe) & \textbf{78.9} & \textbf{74.1} & \textbf{93.7} & 92.7 & 85.4 & \textbf{86.0} & \textbf{60.2} & 29.4 & 23.8 & \textbf{14.0} & \textbf{18.0} & 82.8 & \textbf{72.2} & \textbf{62.4} \\
\midrule
\multicolumn{15}{l}{\textit{Single-flow}} \\[-0.2ex]
QA only & 76.8 & 66.0 & 93.0 & 92.2 & 79.2 & 83.5 & 52.6 & 26.2 & 20.3 & 6.2 & 16.3 & 82.3 & 68.3 & 58.7 \\
Security-analysis only & 74.7 & 69.5 & 92.0 & 90.5 & 83.5 & 82.0 & 55.9 & 22.7 & 19.0 & 7.6 & 14.0 & 83.7 & 69.1 & 58.8 \\
Educational only & 76.7 & 68.2 & 92.9 & 91.8 & 79.1 & 83.8 & 48.7 & \textbf{32.9} & 23.0 & 3.5 & 14.7 & 84.3 & 69.5 & 59.2 \\
Paraphrase only & 76.2 & 65.0 & 92.6 & 90.9 & \textbf{86.3} & 83.2 & 52.4 & 29.3 & 25.3 & 11.8 & 16.8 & 83.7 & 70.6 & 60.3 \\
\midrule
\multicolumn{15}{l}{\textit{Leave-one-out}} \\[-0.2ex]
No QA & 77.8 & 71.5 & 93.1 & 92.2 & 80.6 & 83.8 & 58.2 & 27.7 & 25.0 & 11.6 & 10.0 & \textbf{85.3} & 70.8 & 60.6 \\
No security-analysis & 78.3 & 65.8 & 93.3 & 91.8 & 80.3 & 85.3 & 53.9 & 27.3 & \textbf{26.7} & 5.4 & 11.7 & \textbf{85.3} & 70.2 & 59.6 \\
No educational & 77.7 & 70.4 & 93.3 & 92.9 & 86.2 & 84.4 & 55.3 & 26.9 & 23.0 & 6.4 & \textbf{18.0} & 82.7 & 67.9 & 60.4 \\
No paraphrase & 77.9 & 72.4 & 93.5 & \textbf{93.0} & 82.8 & 84.7 & 60.1 & 29.0 & 23.7 & 9.8 & 15.3 & 83.7 & 67.6 & 61.0 \\
\bottomrule
\end{tabular}
\renewcommand{\arraystretch}{1.0}
\caption{Synthetic flow ablations. All variants share the identical SFT stage and differ only in which synthetic flows compose the mid-training corpus: \textit{single-flow} keeps one flow, \textit{leave-one-out} removes one flow. \textbf{Bold} indicates the best score per column across all variants.}
\label{tab:ablation_flows}
\end{table*}
 
\paragraph{Initialization checkpoint.}
Finally, because no Qwen3-32B-Base checkpoint is publicly available, our 32B pipeline starts from a post-trained checkpoint and is therefore less controlled than the 8B setup. To better understand the effect of starting from a post-trained checkpoint, we run the full 8B pipeline from the post-trained Qwen3-8B and compare it to the original MiST checkpoint, created from Qwen3-8B-Base. As shown in \Cref{tab:ablation_init}, base initialization is slightly but consistently better, both after SFT (62.4 vs.\ 60.6) and after DPO (61.7 vs.\ 60.4). This shows that the base model responds better to our full training methodology.
 
\begin{table*}[!htb]
\centering
\scriptsize
\setlength{\tabcolsep}{2.6pt}
\renewcommand{\arraystretch}{0.92}
\begin{tabular}{@{}l@{\hspace{0.35em}}cccccccccccccc@{}}
\toprule
 & \multicolumn{2}{c}{\textbf{CTI}} & \multicolumn{3}{c}{\textbf{Secure}} & \multicolumn{4}{c}{\textbf{Athena}} & \multicolumn{2}{c}{\textbf{PrimeVul}} &  &  &  \\[-0.5ex]
\cmidrule(lr){2-3} \cmidrule(lr){4-6} \cmidrule(lr){7-10} \cmidrule(lr){11-12} \noalign{\vskip-1.0ex}
\textbf{Model} & MCQ & RCM & MAET & CWET & KCV & CKT & ATE & RMS & TAA & P-C & P-C CoT & \shortstack{MMLU\\Cyber} & \shortstack{Sec\\Eval} & Mean \\
\midrule
MiST-8B-SFT (from Base) & \textbf{78.9} & \textbf{74.1} & \textbf{93.7} & \textbf{92.7} & 85.4 & \textbf{86.0} & \textbf{60.2} & \textbf{29.4} & \textbf{23.8} & \textbf{14.0} & \textbf{18.0} & \textbf{82.8} & \textbf{72.2} & \textbf{62.4} \\
MiST-8B-SFT (from Instruct) & 78.5 & 71.5 & 92.5 & 92.0 & \textbf{85.5} & 85.2 & 55.7 & 28.1 & 22.3 & 13.4 & 11.5 & 81.7 & 70.0 & 60.6 \\
\midrule
MiST-8B-DPO (from Base) & \textbf{79.3} & \textbf{73.9} & \textbf{94.3} & \textbf{93.7} & \textbf{87.3} & \textbf{86.0} & \textbf{57.3} & 25.7 & \textbf{24.6} & 10.0 & \textbf{15.0} & \textbf{83.4} & \textbf{71.5} & \textbf{61.7} \\
MiST-8B-DPO (from Instruct) & 77.4 & 72.2 & 94.1 & 93.4 & 86.4 & 85.3 & 53.9 & \textbf{27.4} & 23.0 & \textbf{12.4} & 10.4 & 80.3 & 69.5 & 60.4 \\
\bottomrule
\end{tabular}
\renewcommand{\arraystretch}{1.0}
\caption{Effect of the initialization checkpoint at the 8B scale. Both pipelines apply the identical mid-training, SFT, and DPO stages and differ only in whether they start from Qwen3-8B-Base or the post-trained Qwen3-8B. \textbf{Bold} indicates the better score per column within each stage.}
\label{tab:ablation_init}
\end{table*}

\section{Cybersecurity Resources and Acronyms}
\label{app:cyber_resources}

\Cref{tab:cyber_resources} summarizes the cybersecurity resources, taxonomies, and
acronyms referenced in the seed corpus and synthetic data generation pipeline. These
resources differ in scope: some provide instance-level vulnerability records, some encode
abstract security concepts and relationships, and others represent operational artifacts
used in detection engineering, threat hunting, incident analysis, and defensive guidance.

\begin{table*}[!htb]
    \centering
    \small
    \setlength{\tabcolsep}{4pt}
    \renewcommand{\arraystretch}{1.0}
        \begin{tabularx}{\textwidth}{@{}
            >{\raggedright\arraybackslash}p{0.13\textwidth}
            >{\raggedright\arraybackslash}p{0.27\textwidth}
            >{\raggedright\arraybackslash}X
        @{}}        
        \toprule
        \textbf{Term} & \textbf{Full name} & \textbf{Description} \\
        \midrule
        NVD
            & National Vulnerability Database
            & A U.S. government vulnerability database that enriches CVE records with metadata such as severity scores, affected products, weakness mappings, and external references. \\
        CVE
            & Common Vulnerabilities and Exposures
            & A standardized identifier system for publicly disclosed cybersecurity vulnerabilities. \\
        CVSS
            & Common Vulnerability Scoring System
            & A scoring framework for representing the severity and exploitability characteristics of vulnerabilities. \\
        CTI
            & Cyber Threat Intelligence
            & Information about threats, adversaries, vulnerabilities, malware, campaigns, tactics, techniques, procedures, and defensive context. \\
        RSS
            & Really Simple Syndication
            & A web feed format used to collect newly published security reports, advisories, and blog posts from external sources. \\
        CWE
            & Common Weakness Enumeration
            & A taxonomy of recurring software and hardware weakness types that can lead to vulnerabilities. \\
        CAPEC
            & Common Attack Pattern Enumeration and Classification
            & A taxonomy of common attack patterns describing how adversaries exploit weaknesses in systems and software. \\
        ATT\&CK
            & Adversarial Tactics, Techniques, and Common Knowledge
            & A knowledge base of adversary tactics, techniques, procedures, software, groups, and campaigns. \\
        D3FEND
            & Digital Artifact and Defensive Knowledge
            & A knowledge graph of defensive cybersecurity techniques, countermeasures, and their relationships to adversary behaviors. \\
        Sigma
            & Sigma rules
            & A generic rule format for describing log-based detection logic in a platform-independent way. \\
        Atomic Red Team
            & Atomic Red Team tests
            & Small adversary-emulation tests designed to exercise specific ATT\&CK techniques in controlled environments. \\
        Splunk ESCU
            & Splunk Enterprise Security Content Update
            & Curated Splunk security content, including detections, analytic stories, and operational guidance for security monitoring. \\
        MISP Galaxy
            & Malware Information Sharing Platform Galaxy
            & A structured vocabulary for representing threat actors, malware families, campaigns, tools, and related threat-intelligence entities. \\
        OWASP
            & Open Worldwide Application Security Project
            & A community-driven source of application-security guidance, risks, testing practices, and secure-development recommendations. \\
        NIST
            & National Institute of Standards and Technology
            & A U.S. standards body that publishes cybersecurity frameworks, controls, guidelines, and technical recommendations. \\
        \bottomrule
    \end{tabularx}
    \caption{Cybersecurity resources, taxonomies, and acronyms referenced in the seed corpus and data generation pipeline.}
    \label{tab:cyber_resources}
\end{table*}
\section{Training Data Composition}
We summarize the composition of the seed corpus and the synthetic training datasets in \Cref{tab:data_stats}. The seed corpus contains raw high-quality cybersecurity sources used to ground data generation. The mid-training split consists primarily of synthetic cybersecurity examples spanning educational explanations, paraphrases, Q\&A, and security-analysis tasks, as explained in \Cref{sec:synth}. The SFT split combines cybersecurity conversations with a broad general-instruction dataset, ensuring that the final training mixture preserves cybersecurity specialization while maintaining general instruction-following ability.

\begin{table*}[!htb]
    \centering
    \small
    \setlength{\tabcolsep}{6pt}
    \renewcommand{\arraystretch}{1.05}
    \begin{tabular}{llrrrc}
        \toprule
        \textbf{Split} & \textbf{Source / flow}               & \textbf{Samples}   & \textbf{Avg tok.} & \textbf{Tokens (M)} & \textbf{Token share (\%)} \\
        \midrule
        \multirow{9}{*}{Seed}
                       & Security reports                      & 3,506              & 2,464             & 8.6                 & --                        \\
                       & CVE records                                  & 156,834            & 152               & 23.9                & --                        \\
                       & MISP Galaxy                           & 8,104              & 299               & 2.4                 & --                        \\
                       & MITRE                                 & 2,182              & 3,205             & 7.0                 & --                        \\
                       & NIST publications                     & 16,827             & 458               & 7.7                 & --                        \\
                       & OWASP                                 & 240                & 3,132             & 0.8                 & --                        \\
                       & CWE examples                          & 940                & 602               & 0.6                 & --                        \\
                       & Security platform guides              & 9,672              & 410               & 4.0                 & --                        \\
                       & Wikipedia security                    & 21,356             & 455               & 9.7                 & --                        \\
        \cmidrule(lr){2-6}
                       & \textit{Seed total}                   & \textit{219,661}   & \textit{--}       & \textit{64.7}       & \textit{--}               \\
        \midrule
        \multirow{5}{*}{Mid-training}
                       & Cybersecurity: Educational            & 254,287            & 2,416             & 614.5               & 63.0                      \\
                       & Cybersecurity: Paraphrases            & 233,828            & 780               & 182.4               & 18.7                      \\
                       & Cybersecurity: Q\&A                   & 576,087            & 115               & 66.1                & 6.8                       \\
                       & Cybersecurity: Security Analysis      & 123,743            & 589               & 72.9                & 7.5                       \\
                       & Dolci: General instruction subset      & 50,000             & 798               & 39.9                & 4.1                       \\
        \cmidrule(lr){2-6}
                       & \textit{Mid-training total}           & \textit{1,237,945} & \textit{--}       & \textit{975.7}      & \textit{100.0}            \\
        \midrule
        \multirow{2}{*}{SFT}
                       & Cybersecurity: Conversations          & 307,679            & 1,145             & 352.2               & 22.7                      \\
                       & Dolci: General instruction (full)     & 1,924,533          & 623               & 1,198.7             & 77.3                      \\
        \cmidrule(lr){2-6}
                       & \textit{SFT total}                    & \textit{2,232,212} & \textit{--}       & \textit{1,550.9}    & \textit{100.0}            \\
        \midrule
        \multirow{3}{*}{Total}
                       & Cybersecurity                         & 1,495,624          & --                & 1,288.0             & 51.0                      \\
                       & General Instructions                  & 1,974,533          & --                & 1,238.6             & 49.0                      \\
                       & {Training total (excl.\ seed)}        & {3,470,157}        & {--}              & {2,526.6}           & 100.0                     \\
        \bottomrule
    \end{tabular}
    \caption{Data statistics for the curated seed corpus and the synthetic datasets used for mid-training and SFT. Seed token counts are computed using the Qwen3 tokenizer. Token counts are reported in millions (M), and Avg tok.\ denotes the mean number of tokens per sample. Token share (\%) is shown only for Mid-training and SFT (computed within each split), and for Total (computed w.r.t.\ Training total (excl.\ seed)).}
    \label{tab:data_stats}
\end{table*}

\section{Model Identifiers and Hugging Face URLs}
\label{app:model_urls}

For reproducibility, \Cref{tab:model_urls} maps the open model names used in our
result tables to their corresponding Hugging Face repositories when available. Proprietary
OpenAI models reported in the tables, including GPT-5.4-mini and GPT-5.4-nano, are
not included because they are not released as Hugging Face model checkpoints.

\begin{table*}[!htb]
    \centering
    \small
    \setlength{\tabcolsep}{5pt}
    \renewcommand{\arraystretch}{1.08}
    \begin{tabular}{ll}
        \toprule
        \textbf{Name in tables} & \textbf{Hugging Face repository} \\
        \midrule
        Qwen3-235B
            & \href{https://huggingface.co/Qwen/Qwen3-235B-A22B-Instruct-2507}
            {\texttt{Qwen/Qwen3-235B-A22B-Instruct-2507}} \\
        DeepHat-7B
            & \href{https://huggingface.co/DeepHat/DeepHat-V1-7B}
            {\texttt{DeepHat/DeepHat-V1-7B}} \\
        Foundation-Sec-8B
            & \href{https://huggingface.co/fdtn-ai/Foundation-Sec-8B-Instruct}
            {\texttt{fdtn-ai/Foundation-Sec-8B-Instruct}} \\
        Lily-Cyber-7B
            & \href{https://huggingface.co/segolilylabs/Lily-Cybersecurity-7B-v0.2}
            {\texttt{segolilylabs/Lily-Cybersecurity-7B-v0.2}} \\
        Primus-70B
            & \href{https://huggingface.co/trend-cybertron/Llama-Primus-Nemotron-70B-Instruct}
            {\texttt{trend-cybertron/Llama-Primus-Nemotron-70B-Instruct}} \\
        Primus-8B
            & \href{https://huggingface.co/trend-cybertron/Llama-Primus-Merged}
            {\texttt{trend-cybertron/Llama-Primus-Merged}} \\
        RedSage-8B-DPO
            & \href{https://huggingface.co/RISys-Lab/RedSage-Qwen3-8B-DPO}
            {\texttt{RISys-Lab/RedSage-Qwen3-8B-DPO}} \\
            Qwen3-8B-Base
    & \href{https://huggingface.co/Qwen/Qwen3-8B-Base}
    {\texttt{Qwen/Qwen3-8B-Base}} \\
        Qwen3-8B
            & \href{https://huggingface.co/Qwen/Qwen3-8B}
            {\texttt{Qwen/Qwen3-8B}} \\
        Qwen3-32B
            & \href{https://huggingface.co/Qwen/Qwen3-32B}
            {\texttt{Qwen/Qwen3-32B}} \\
        CyberPal2.0-20B
            & \href{https://huggingface.co/cyber-pal-security/CyberPal2.0-20B}
            {\texttt{cyber-pal-security/CyberPal2.0-20B}} \\
        \bottomrule
    \end{tabular}
    \caption{Mapping between the open model names reported in the paper tables and their corresponding Hugging Face repositories when available. Names are shown as they appear in the result tables.}
    \label{tab:model_urls}
\end{table*}

\section{Decontamination}
\label{app:decontamination}

We employ a multi-layered decontamination pipeline that operates at three different points
in the data lifecycle: (i) source-level filtering of the seed corpus, (ii) structural
constraints during synthetic data generation, and (iii) $n$-gram overlap filtering applied
to every training stage (mid-training, SFT, and DPO). This combination is designed to
prevent both \emph{identifier leakage} (specific CVE, CWE, or commit-hash entities that
appear in our benchmarks) and \emph{textual leakage} (verbatim spans copied into training
examples), following established protocols~\citep{brown2020language,dodge2021documenting}.
Decontamination matters because prior work has shown that pretraining-time exposure to
benchmark instances can be exploited at evaluation time, inflating reported
performance~\citep{magar2022data,sainz2023nlp}.

\paragraph{Source-level filtering.}
Before the seed corpus is used for any downstream transformation, we audit each source
against the evaluation benchmarks and remove records that anchor a benchmark example,
following the practice of removing benchmark-related source data adopted in recent
domain-adapted LLMs~\citep{hui2024qwen2,jacovi2023stop}. We remove all CVE records that appear in
the CTI-Bench root-cause-mapping (\texttt{cti-rcm}) and vulnerability-severity-prediction
(\texttt{cti-vsp}) splits from the CVE seed corpus before running any synthetic generation
flow. For the GHSA-linked vulnerable and fixed code records, we additionally remove any
record whose fixing commit hash overlaps with the PrimeVul (test and validation splits); code contamination has been shown to substantially inflate
performance on code benchmarks such as HumanEval and MBPP~\citep{riddell2024quantifying},
which motivates this stricter, identifier-level filtering for the code portion of our
seed. We log the number of removed records and the offending CVE identifiers per benchmark
so that the audit is reproducible.

\paragraph{Format-level constraints on synthetic data.}
Because the majority of our evaluation benchmarks are multiple-choice (MCQ), we
deliberately do \emph{not} include MCQ-style generation prompts in any synthetic flow
described in \Cref{sec:synth}. This avoids exposing the model to the structural and
stylistic patterns of MCQs during mid-training and SFT (e.g., the ``A/B/C/D'' option
layout and stem--distractor phrasing), which have been shown to constitute an indirect
contamination vector: \citet{deng2024investigating} demonstrate that models can recover
masked \emph{incorrect} options on MMLU at high exact-match rates, indicating that exposure
to MCQ structure itself leaks evaluation signal. Our synthetic flows therefore produce
free-form paraphrases, educational rewrites, short open-ended QA, conversations, and
persona-based dialogues, none of which mirror MCQ format.

\paragraph{$N$-gram overlap filtering.}
We adopt a 13-gram overlap criterion~\citep{brown2020language} as the primary mechanism
for detecting textual contamination across all training data. Token- and word-level
$n$-gram matching is the standard decontamination tool used in modern LLM training,
including Llama~2~\citep{touvron2023llama2}, and
Qwen2.5-Coder~\citep{hui2024qwen2}. We construct the decontamination corpus by aggregating,
for every benchmark in our cybersecurity evaluation suite (\Cref{sec:cyber_benchmarks}),
both the question prompts and the reference answers, including the per-task instruction
templates when applicable. For multi-turn or message-style benchmarks, each message content
string is included independently. All text is normalized by lower-casing and stripping all
non-alphanumeric characters before $n$-gram extraction, following the normalization
recommended by~\citet{dodge2021documenting}, which makes the matching robust to differences
in punctuation, whitespace, and casing between benchmark and training text. We extract
overlapping $n$-grams using a sliding window with $n = 13$ and store the unique benchmark
$n$-grams in a Python hash set for constant-time lookup. A training sample is flagged as
contaminated as soon as a single matching $n$-gram is found (threshold $= 1$), and
matching is short-circuited on the first hit for efficiency.

\paragraph{Pipeline integration.}
The $n$-gram filter is applied as a preprocessing step inside the training entry point,
immediately after dataset loading and instruction formatting and before tokenization, so
that the model never sees contaminated samples in any stage. The same filter, configured
identically, is applied at mid-training, SFT, and DPO. For each run, we serialize each
sample into a single text by concatenating the relevant text fields (e.g., \texttt{messages},
\texttt{prompt}, \texttt{text}, \texttt{instruction}, \texttt{input}, \texttt{output}); for
message-formatted data we concatenate the contents of all turns so that both user and
assistant tokens are checked. The benchmark $n$-gram set is cached to disk after the first
construction to amortize cost across runs, and a per-run contamination report is written
to the model output directory, including the total number of samples, the number and
fraction flagged as contaminated, the configured $n$-gram size and threshold, and a sample
of contaminated examples (with their matching $n$-grams and text previews) for manual
inspection. This per-run reporting aligns with the per-benchmark contamination
documentation called for by~\citet{sainz2023nlp}.

\paragraph{Statistics.}
Across the cybersecurity benchmarks listed in \Cref{tab:general-benchmarks} and \Cref{tab:cyber-benchmarks}, the aggregated decontamination corpus comprises $27{,}201$ benchmark text
segments yielding approximately $467{,}000$ unique 13-grams. Applying the full pipeline to
the combined mid-training and SFT corpus described in \Cref{tab:data_stats} removes
$\approx 3.78\%$ of the candidate training samples, with the majority of removals
originating from synthetic flows grounded in MITRE and CVE seeds whose surface form
occasionally overlaps with benchmark prompts that reference the same identifiers.

\paragraph{Limitations of $n$-gram-based decontamination.}
$N$-gram overlap is robust to verbatim and lightly modified reuse, but is known to miss
heavily paraphrased or translated variants of benchmark items~\citep{yang2024rethinking};
black-box statistical tests~\citep{oren2024proving} provide complementary evidence but
cannot replace source-level removal. Our reliance on source-level filtering for
benchmark-anchoring identifiers (CVE records, commit hashes) and on the deliberate exclusion of
MCQ-style synthetic generation is intended to mitigate these residual risks where they are
most likely to matter for cybersecurity evaluation.

\section{RSS Blogposts Curation}\label{app:blog_filter}
As described in \Cref{sec:seed}, we include cybersecurity blog posts from RSS feeds as part
of the seed corpus to complement structured security knowledge bases with timely,
practitioner-driven insights. These posts provide unstructured, narrative accounts of
real-world incidents, attack techniques, vulnerabilities, and defensive practices. However,
many entries contain broadly scoped security commentary without substantive technical
content, or are written in languages other than English, necessitating additional filtering
to ensure domain relevance and quality.

 To ensure corpus quality and domain relevance, we apply an automated filtering pipeline in which an LLM (GPT-5-Mini) acts as a judge to identify and exclude posts that are not directly relevant to cyber threat intelligence (CTI). While our broader focus is on cybersecurity, we restrict this subset to CTI-oriented content in order to retain posts that provide concrete, actionable descriptions of threats, attacker behavior, and real-world incidents, rather than high-level or purely defensive commentary. The full prompt used for filtering is shown in \Cref{fig:filter_prompt}. Applying this pipeline reduces the number of blog posts from $7{,}916$ to $2{,}995$.

\begin{figure}[!htb]
    \centering
    \begin{tcolorbox}[
            title=Relevance Filtering Prompt,
            colback=white,
            colframe=black,
            coltitle=white,
            fonttitle=\bfseries,
            boxrule=0.9pt,
            arc=1.2mm,
            breakable,
            left=6pt,right=6pt,top=6pt,bottom=6pt,
            before skip=10pt, after skip=10pt
        ]
        \footnotesize
        \setlength{\parindent}{0pt}
        \setlength{\parskip}{4pt}

        \textbf{Task.} Score the report on two dimensions:

        \begin{enumerate}
            \item \textbf{english\_score} (1--10): How much of the report is written in English.
                  \begin{itemize}
                      \item 1 = Not in English at all (completely in another language)
                      \item 5 = Mixed languages (partially in English)
                      \item 10 = Fully in English
                  \end{itemize}

            \item \textbf{cti\_score} (1--10): How related the report is to Cyber Threat Intelligence (CTI).
                  \begin{itemize}
                      \item 1 = Not related to CTI
                      \item 5 = Security-related but not CTI-specific
                      \item 10 = Strong CTI focus (threat actors, malware, vulnerabilities, techniques, IOCs, etc.)
                  \end{itemize}
        \end{enumerate}

        \textbf{Report:}
        \begin{tcolorbox}[
                colback=white,
                colframe=black!40,
                boxrule=0.6pt,
                arc=1mm,
                left=5pt,right=5pt,top=4pt,bottom=4pt
            ]
            \ttfamily\footnotesize
            {report\_content}
        \end{tcolorbox}

        \textbf{Output.} Return \emph{only} valid JSON (no markdown, no extra text):
        \begin{tcolorbox}[
                colback=white,
                colframe=black!40,
                boxrule=0.6pt,
                arc=1mm,
                left=5pt,right=5pt,top=4pt,bottom=4pt
            ]
            \ttfamily\footnotesize
            \{"english\_score": <number>, "cti\_score": <number>\}
        \end{tcolorbox}

    \end{tcolorbox}
    \caption{Scoring instructions for RSS feed filtering.}
    \label{fig:filter_prompt}
\end{figure}

\section{Task-Specific Reinforcement Learning Details}\label{app:rl}

We apply Group Relative Policy Optimization (GRPO;~\citealp{shao2024deepseekmath}) to MiST-8B and the Qwen3-8B baseline on three verifiable cybersecurity tasks. Each task admits deterministic, rule-based reward functions that require no learned reward model.

\paragraph{CVE-to-CWE mapping.} Given a CVE description, the model predicts the corresponding CWE identifier. Reward is 1 if the predicted CWE matches the ground-truth label and 0 otherwise.

\paragraph{CVE-to-CVSS vector scoring.} Given a CVE description, the model produces a CVSS vector string. Reward is computed as the fraction of CVSS metric components that match the reference vector.

\paragraph{Vulnerable-code-to-CWE mapping.} Given a code snippet containing a known vulnerability, the model predicts the associated CWE. Scoring follows the same exact-match criterion as CVE-to-CWE mapping.

\section{Task-Specific SFT Details}
\label{app:primevul_sft}

For supervised task adaptation, we fine-tune the Qwen and MiST checkpoints on
PrimeVul~\citep{primevul}. PrimeVul is a vulnerability-detection benchmark constructed
from real-world C/C++ functions and includes paired examples in which a vulnerable
function is matched with its corresponding patched version. This paired structure makes it
possible to evaluate whether a model can identify the security-relevant difference between
two closely related functions.

Training uses only the PrimeVul training and validation splits. Each example is converted
into a three-turn chat format consisting of: (i) a system message instructing the model to
act as a security expert, (ii) a user message containing the function and a direct
YES/NO vulnerability-detection instruction, and (iii) an assistant response containing the
gold label. The paired subset is used as provided by PrimeVul, while the default subset is
class-balanced within each split.

We evaluate on the held-out PrimeVul paired test split using two prompting variants. The
first, denoted \textsc{PrimeVul}, matches the direct YES/NO format used during SFT. The
second, denoted \textsc{PrimeVul-CoT}, asks the model to reason step-by-step before
producing its final verdict. In both cases, the final answer is extracted as a binary
YES/NO prediction.

Following the PrimeVul evaluation protocol, we report paired accuracy rather than standard
accuracy. Standard accuracy can be misleading in vulnerability detection because models may
learn dataset priors or over-predict the vulnerable class without identifying the precise
code-level distinction between a vulnerable function and its patched counterpart. Under
paired accuracy, a pair is counted as correct only if both the vulnerable function and its
patched version are classified correctly.

The full PrimeVul SFT results are shown in \Cref{tab:primevul_sft}. Before task-specific
SFT, MiST already outperforms the corresponding Qwen baselines under paired accuracy,
suggesting that cybersecurity mid-training improves sensitivity to vulnerability-relevant
signals. After task-specific SFT, the gap becomes larger: MiST-8B-DPO-PrimeVul improves
from 10.3 to 16.3 on \textsc{PrimeVul} and from 15.2 to 19.2 on
\textsc{PrimeVul-CoT}, while MiST-32B-DPO-PrimeVul improves from 8.6 to 16.4 and from
15.3 to 21.3, respectively. In contrast, Qwen gains only marginally after the same
task-specific SFT procedure. This indicates that cybersecurity mid-training provides a
better initialization for learning the paired vulnerability-detection objective.

\begin{table}[!hbt]
\centering
\small
\begin{tabular}{lcc}
\toprule
\textbf{Model} & P-C & P-C CoT \\
\midrule
\multicolumn{3}{l}{\textit{PrimeVul-8B}} \\
Qwen3-8B-Instruct & 2.2 & 13.6 \\
Qwen3-8B-PrimeVul & 2.8 & 13.5 \\
MiST-8B-DPO & 10.3 & 15.2 \\
MiST-8B-DPO-PrimeVul & 16.3 & 19.2 \\
\midrule
\multicolumn{3}{l}{\textit{PrimeVul-32B}} \\
Qwen3-32B-Instruct & 4.7 & 16.4 \\
Qwen3-32B-PrimeVul & 5.2 & 18.2 \\
MiST-32B-DPO & 8.6 & 15.3 \\
MiST-32B-DPO-PrimeVul & 16.4 & 21.3 \\
\bottomrule
\end{tabular}
\caption{Task-specific SFT on PrimeVul. We report paired accuracy (P-C), following PrimeVul's
pair-wise evaluation protocol.}
\label{tab:primevul_sft}
\end{table}

\section{Data examples}\label{app:examples}
We present representative examples of synthetic data generated by our pipeline described in
\Cref{sec:cyber_midtraining_data}, including persona-conditioned dialogues
(\Cref{fig:personas}), standalone QA pairs (\Cref{fig:qa}), and educational content
(\Cref{fig:personas}).

\section{Safety Evaluation}
\label{app:safety}

Because cybersecurity specialization carries dual-use risk, we ran a
preliminary safety evaluation of the MiST 8B checkpoints, focused on whether
task specialization degrades the model's tendency to refuse or safely handle
misuse-oriented requests relative to its base model.

\paragraph{Setup.}
We evaluated on the CyberSecEval MITRE benchmark~\cite{wan2024cyberseceval3}, which
probes willingness to assist with offensive-security tasks mapped to the
MITRE ATT\&CK framework. Rather than the benchmark's original judge, we scored responses with the Llama-3.3 judge of \citet{liu2025purpcode}, which they introduce to reduce false positives from CyberSecEval MITRE's original judge. We compare the safe-response rates of MiST-8B-SFT and MiST-8B-DPO against the Qwen3-8B base model.

\paragraph{Results.}
Both MiST checkpoints improve over the Qwen3-8B baseline in safe-response
rate (Table~\ref{tab:safety}), indicating that specialization did not erode
safe-handling behavior on this benchmark.

\begin{table}[h]
\centering
\begin{tabular}{lc}
\toprule
Model & Safe-response rate (\%) \\
\midrule
Qwen3-8B (base) & 75.8 \\
MiST-8B-SFT     & 89.8 \\
MiST-8B-DPO     & 85.3 \\
\bottomrule
\end{tabular}
\caption{Safe-response rate on the CyberSecEval MITRE benchmark}
\label{tab:safety}
\end{table}

\paragraph{Qualitative red-teaming and internal benchmark.}
We complemented the automated evaluation with qualitative red-teaming and an
internal safety benchmark covering misuse-oriented prompts. Across these,
MiST performed comparably to or better than the baseline.

\clearpage
\begin{figure*}[!htb]
\centering
\begin{tcolorbox}[
  title=Conversation with a Blue Team Defender,
  colback=gray!3,
  colframe=black!80,
  coltitle=white,
  fonttitle=\bfseries,
  boxrule=0.9pt,
  arc=1.2mm,
  breakable,
  left=6pt,right=6pt,top=6pt,bottom=6pt,
  before skip=10pt, after skip=10pt
]
\footnotesize
\setlength{\parindent}{0pt}
\setlength{\parskip}{4pt}
\begin{tcolorbox}[title=Persona]
\small
\textbf{Role:} Blue Team Defender (Endpoint Security Analyst). \\
\textbf{Goal:} Understand how attackers abuse shell configuration files for persistence and derive practical EDR/XDR detection rules for unauthorized writes (e.g., \texttt{/etc/profile}, \textasciitilde{}/\texttt{.zshrc}).
\end{tcolorbox}
\begin{tcolorbox}[
  title=User,
  colback=cyan!12,
  colframe=cyan!60,
  boxrule=0.6pt,
  arc=1mm,
  left=5pt,right=5pt,top=4pt,bottom=4pt,
  breakable
]
\small
I've been reviewing some EDR alerts and noticed a few instances where a User's \textasciitilde{}/.bash\_profile was modified by a non-root process. Since this file is typically only writable by root, I'm concerned this could be a persistence attempt. How should I interpret this, and what should I look for next?
\end{tcolorbox}

\begin{tcolorbox}[
  title=Assistant,
  colback=orange!10,
  colframe=orange!60,
  boxrule=0.6pt,
  arc=1mm,
  left=5pt,right=5pt,top=4pt,bottom=4pt,
  breakable
]
\small
That’s a solid red flag. Modifying \textasciitilde{}/.bash\_profile by a non-root process is highly suspicious and aligns with the MITRE ATT\&CK technique **Event Triggered Execution: Unix Shell Configuration Modification (T1546.004)**. Adversaries often exploit weak file permissions or privilege escalation to write malicious commands into User-level shell config files. You should immediately check the process responsible for the write—look for unusual parent processes, unexpected command-line arguments, or execution from temporary directories. Also, verify if the modification includes commands that launch binaries, download payloads, or establish reverse shells.
\end{tcolorbox}

\begin{tcolorbox}[
  title=User,
  colback=cyan!12,
  colframe=cyan!60,
  boxrule=0.6pt,
  arc=1mm,
  left=5pt,right=5pt,top=4pt,bottom=4pt,
  breakable
]
\small
Good point. I found the process was a Python script running from /tmp with no User interaction. The change added a line to source a script from a remote URL. Is this a common pattern in known malware?
\end{tcolorbox}

\begin{tcolorbox}[
  title=Assistant,
  colback=orange!10,
  colframe=orange!60,
  boxrule=0.6pt,
  arc=1mm,
  left=5pt,right=5pt,top=4pt,bottom=4pt,
  breakable
]
\small
Yes, exactly. This is a classic tactic seen in several campaigns, including those tied to the Rocke threat actor. The use of a Python script from /tmp to modify \textasciitilde{}/.bash\_profile and source a remote payload fits the **T1546.004** pattern. The script may be designed to execute on every new login, ensuring persistence. You should also check for similar modifications in other shell config files like \textasciitilde{}/.bashrc or \textasciitilde{}/.profile—adversaries often use multiple files to increase reliability, especially if one is missing or ignored.
\end{tcolorbox}

\begin{tcolorbox}[
  title=User,
  colback=cyan!12,
  colframe=cyan!60,
  boxrule=0.6pt,
  arc=1mm,
  left=5pt,right=5pt,top=4pt,bottom=4pt,
  breakable
]
\small
Got it. I’ve seen similar behavior on macOS systems too. How does this differ on macOS, especially since zsh is now the default shell?
\end{tcolorbox}

\begin{tcolorbox}[
  title=Assistant,
  colback=orange!10,
  colframe=orange!60,
  boxrule=0.6pt,
  arc=1mm,
  left=5pt,right=5pt,top=4pt,bottom=4pt,
  breakable
]
\small
On macOS, the behavior is similar but uses zsh configuration files. The login shell sources /etc/zprofile, /etc/zshenv, and \textasciitilde{}/.zprofile, while the interactive shell uses \textasciitilde{}/.zshrc. Malware has been observed modifying \textasciitilde{}/.zshrc or \textasciitilde{}/.zprofile to execute malicious commands on login. In fact, some macOS-specific malware like PersistentJXA leverages zsh config files to maintain persistence. You should monitor writes to \textasciitilde{}/.zshrc, \textasciitilde{}/.zprofile, and \textasciitilde{}/.zlogin—especially from non-privileged processes. Also, note that macOS may still load /etc/bashrc for legacy compatibility, so keep an eye on that too.
\end{tcolorbox}

\begin{tcolorbox}[
  title=User,
  colback=cyan!12,
  colframe=cyan!60,
  boxrule=0.6pt,
  arc=1mm,
  left=5pt,right=5pt,top=4pt,bottom=4pt,
  breakable
]
\small
That was helpful, thanks!
\end{tcolorbox}

\begin{tcolorbox}[
  title=Assistant,
  colback=orange!10,
  colframe=orange!60,
  boxrule=0.6pt,
  arc=1mm,
  left=5pt,right=5pt,top=4pt,bottom=4pt,
  breakable
]
\small
You’re very welcome!
If you have any other questions, please feel free to ask.
\end{tcolorbox}

\end{tcolorbox}
\caption{Example of persona-conditioned dialogue. The persona casts the \emph{user} as a Blue Team defender, steering the interaction toward defensive investigation. The assistant provides expert guidance on identifying persistence mechanisms via shell configuration changes on Linux and macOS.}
\label{fig:personas}
\end{figure*}

\clearpage

\begin{figure*}[!htb]
\centering
\begin{tcolorbox}[
  title=QA 1: MFA Integration for Legacy Systems,
  colback=gray!3,
  colframe=black!80,
  coltitle=white,
  fonttitle=\bfseries,
  boxrule=0.9pt,
  arc=1.2mm,
  breakable,
  left=6pt,right=6pt,top=6pt,bottom=6pt,
  before skip=10pt, after skip=10pt
]
\footnotesize
\setlength{\parindent}{0pt}
\setlength{\parskip}{4pt}

\begin{tcolorbox}[
  title=User,
  colback=cyan!12,
  colframe=cyan!60,
  boxrule=0.6pt,
  arc=1mm,
  left=5pt,right=5pt,top=4pt,bottom=4pt,
  breakable
]
\small
How do third-party tools like Duo Security or Thales SafeNet enable Multi-factor Authentication (MFA) for legacy systems that lack native MFA support, and what type of authentication protocol is typically used to integrate these tools with older infrastructure?
\end{tcolorbox}

\begin{tcolorbox}[
  title=Assistant,
  colback=orange!10,
  colframe=orange!60,
  boxrule=0.6pt,
  arc=1mm,
  left=5pt,right=5pt,top=4pt,bottom=4pt,
  breakable
]
\small
Third-party tools like Duo Security and Thales SafeNet enable MFA for legacy systems by acting as an intermediary authentication layer. They integrate with older systems through standard protocols such as RADIUS (Remote Authentication Dial-In User Service) or TACACS+ (Terminal Access Controller Access-Control System Plus). These protocols allow the legacy system to forward authentication requests to the MFA provider, which then verifies the User via a second factor (e.g., push notification, one-time password). This enables MFA enforcement without requiring changes to the legacy application or infrastructure.
\end{tcolorbox}
\end{tcolorbox}

\begin{tcolorbox}[
  title=QA 2: OAuth Abuse in Microsoft 365,
  colback=gray!3,
  colframe=black!80,
  coltitle=white,
  fonttitle=\bfseries,
  boxrule=0.9pt,
  arc=1.2mm,
  breakable,
  left=6pt,right=6pt,top=6pt,bottom=6pt,
  before skip=10pt, after skip=10pt
]
\footnotesize
\setlength{\parindent}{0pt}
\setlength{\parskip}{4pt}

\begin{tcolorbox}[
  title=User,
  colback=cyan!12,
  colframe=cyan!60,
  boxrule=0.6pt,
  arc=1mm,
  left=5pt,right=5pt,top=4pt,bottom=4pt,
  breakable
]
\small
What role does a service principal play in enabling a malicious OAuth application to access resources in a Microsoft 365 environment, and why is it necessary to create one?
\end{tcolorbox}

\begin{tcolorbox}[
  title=Assistant,
  colback=orange!10,
  colframe=orange!60,
  boxrule=0.6pt,
  arc=1mm,
  left=5pt,right=5pt,top=4pt,bottom=4pt,
  breakable
]
\small
In Microsoft 365, a service principal acts as the identity for an application within Azure Active Directory. When a malicious OAuth application is created, it must be associated with a service principal to be able to authenticate and access resources. Without a service principal, the application has no identity in the directory and cannot request or receive access tokens. Therefore, creating a service principal is a necessary step for the application to function and gain access to data on behalf of Users, even if the original User is disabled.
\end{tcolorbox}
\end{tcolorbox}

\begin{tcolorbox}[
  title=QA 3: Local-only Indicator Removal,
  colback=gray!3,
  colframe=black!80,
  coltitle=white,
  fonttitle=\bfseries,
  boxrule=0.9pt,
  arc=1.2mm,
  breakable,
  left=6pt,right=6pt,top=6pt,bottom=6pt,
  before skip=10pt, after skip=10pt
]
\footnotesize
\setlength{\parindent}{0pt}
\setlength{\parskip}{4pt}

\begin{tcolorbox}[
  title=User,
  colback=cyan!12,
  colframe=cyan!60,
  boxrule=0.6pt,
  arc=1mm,
  left=5pt,right=5pt,top=4pt,bottom=4pt,
  breakable
]
\small
What is the significance of the "supports\_remote\_execution: False" field in the context of Indicator Removal on Host: Uninstall Malicious Application (T1630.001), and how does it affect the adversary's operational approach?
\end{tcolorbox}

\begin{tcolorbox}[
  title=Assistant,
  colback=orange!10,
  colframe=orange!60,
  boxrule=0.6pt,
  arc=1mm,
  left=5pt,right=5pt,top=4pt,bottom=4pt,
  breakable
]
\small
This field indicates that the technique cannot be executed remotely; it must be performed directly on the compromised device. As a result, the adversary must already have local access or execute the uninstallation through a previously established local foothold, limiting the technique’s use in large-scale, automated attacks and requiring manual or local execution steps.
\end{tcolorbox}
\end{tcolorbox}
\caption{{Representative question–answer pairs used in our synthetic data pipeline.} Each example targets a distinct cybersecurity knowledge area, including enterprise authentication mechanisms, cloud identity abuse, and MITRE ATT\&CK technique semantics, and is designed to elicit concise, technically grounded explanations.}
\label{fig:qa}
\end{figure*}

\clearpage
\begin{figure*}[!htb]
\centering
\begin{tcolorbox}[
  title=Obfuscated Files or Information: Steganography (T1406.001),
  colback=gray!3,
  colframe=black!80,
  coltitle=white,
  fonttitle=\bfseries,
  boxrule=0.9pt,
  arc=1.2mm,
  breakable,
  left=6pt,right=6pt,top=6pt,bottom=6pt,
  before skip=10pt, after skip=10pt
]
\small
\setlength{\parindent}{0pt}
\setlength{\parskip}{5pt}

\textbf{Overview.}
Steganography is a defense evasion technique that conceals data within benign digital media so that the presence of the hidden information is not apparent to users or security controls. In the MITRE ATT\&CK framework, this behavior is classified as \textbf{T1406.001}, a sub-technique of Obfuscated Files or Information. Unlike encryption, which clearly signals protected content, steganography embeds payloads within ordinary files such as images, audio, or text, enabling stealthy storage, delivery, and exfiltration of data. This technique is frequently observed in mobile threat activity, particularly on Android.

\textbf{Mechanisms.}
Steganography exploits redundancy or perceptual limits in common file formats. Typical implementations include:
\begin{itemize}
  \item \textbf{Image-based:} Least significant bit (LSB) modification in BMP and PNG files, or manipulation of discrete cosine transform (DCT) coefficients in JPEG images, allowing data to be hidden without visible distortion.
  \item \textbf{Audio-based:} LSB embedding in uncompressed WAV files and psychoacoustic masking in compressed formats such as MP3, where changes fall below human auditory perception.
  \item \textbf{Text-based:} Encoding information through whitespace variation, controlled word selection, or insertion of zero-width Unicode characters invisible to readers.
\end{itemize}

Embedded data is commonly encrypted prior to insertion, increasing resistance to inspection even if the carrier is suspected.

\textbf{Threat Implications and Use Cases.}

Adversaries leverage steganography to support multiple stages of the attack lifecycle, including second-stage payload delivery, covert storage of configuration or command-and-control data, and stealthy data exfiltration via trusted file-sharing or messaging channels. On Android, this technique is particularly effective due to shared external storage, high volumes of user-generated media, and frequent transmission of images and audio through social and cloud-based applications. Malicious apps can extract hidden content from media files without elevated privileges, bypassing many static and network-based defenses.

\textbf{Detection and Mitigation.}

Detection is challenging because steganographic files typically lack clear indicators of compromise and exhibit statistically natural data distributions. Effective defensive measures include:
\begin{itemize}
  \item Behavioral monitoring for anomalous access to large numbers of media files
  \item File integrity monitoring on shared and external storage locations
  \item Restrictive enforcement of media and storage-related permissions
  \item Targeted use of statistical steganalysis during forensic or high-risk investigations
\end{itemize}

\textbf{Conclusion.}

Steganography under T1406.001 represents a subtle yet effective method for defense evasion, persistence, and covert data movement. Its reliance on trusted media formats enables adversaries to bypass many conventional detection mechanisms, particularly in mobile environments. Addressing this threat requires a shift from signature-based controls toward behavioral analysis and selective forensic inspection of high-risk media content.

\textbf{References.}

MITRE ATT\&CK Framework: Obfuscated Files or Information – Steganography (T1406.001)
\end{tcolorbox}
\caption{{Example of educational content used in our synthetic data pipeline.} The passage provides a technically grounded overview of a MITRE ATT\&CK technique (T1406.001), including mechanisms, threat implications, and detection considerations, reflecting the style of explanatory artifacts generated to support knowledge acquisition.}
\label{fig:education}
\end{figure*}
\clearpage

\end{document}